\documentclass[12pt,preprint]{emulateapj}
\usepackage{hyperref}

\shorttitle{Ultra-deep $J$ and $K_S$ imaging in the ECDFS}
\shortauthors{Hsieh et al.}

\begin{document}

\title{The Taiwan ECDFS Near-Infrared Survey:
Ultra-deep $J$ and $K_S$ Imaging in the Extended Chandra Deep Field-South} 

\author{Bau-Ching Hsieh\altaffilmark{1}, Wei-Hao Wang\altaffilmark{1},
Chih-Chiang Hsieh\altaffilmark{1,2},
Lihwai Lin\altaffilmark{1},
Haojing Yan\altaffilmark{3}, 
Jeremy Lim\altaffilmark{1,4}, Paul Ho\altaffilmark{1,5}
}

\altaffiltext{1}{Institute of Astrophysics \& Astronomy, Academia Sinica,
P.O. Box 23-141, Taipei 106, Taiwan, R.O.C.}

\altaffiltext{2}{Institute of Astronomy, National Tsing Hua University,
No. 101, Section 2, Kuang-Fu Road, Hsinchu, Taiwan 30013, R.O.C}

\altaffiltext{3}{Department of Physics and Astronomy,
University of Missouri, Columbia, MO 65211, USA}

\altaffiltext{4}{Department of Physics, 
University of Hong Kong, Pokfulam Road, Hong Kong}

\altaffiltext{5}{Harvard-Smithsonian Center for Astrophysics, 
60 Garden Street, Cambridge, MA 02138, USA}

\begin{abstract}
We present ultra-deep $J$ and $K_S$ imaging observations
covering a $30'\times30'$ area
of the Extended Chandra Deep Field-South (ECDFS)
carried out by our Taiwan ECDFS Near-Infrared Survey (TENIS).
The median $5\sigma$ limiting magnitudes 
for all detected objects in the ECDFS 
reach 24.5 and 23.9 mag (AB) for $J$ and $K_S$, respectively.
In the inner 400 arcmin$^2$ region where the sensitivity is more uniform, 
objects as faint as 25.6 and 25.0 mag are detected at $5\sigma$. 
So this is by far the
deepest $J$ and $K_S$ datasets available for the ECDFS.
To combine the TENIS with the $Spitzer$ IRAC data
for obtaining better spectral energy distributions of high-redshift objects,
we developed a novel deconvolution technique (IRACLEAN)
to accurately estimate the IRAC fluxes.
IRACLEAN can minimize the effect of blending
in the IRAC images caused by the large point-spread functions
and reduce the confusion noise.
We applied IRACLEAN to the images from 
the $Spitzer$ IRAC/MUSYC Public Legacy in the ECDFS survey (SIMPLE) 
and generated a $J$+$K_S$ selected multi-wavelength catalog
including the photometry of both the TENIS near-infrared 
and the SIMPLE IRAC data.
We publicly release the data products derived from this work, including the
$J$ and $K_S$ images and the $J$+$K_S$ selected multiwavelength catalog.
\end{abstract}


\section{Introduction}\label{introduction}
Near-Infrared (NIR) imaging surveys provide 
several advantages over optical observations for studies of galaxies.
For nearby galaxies, 
stellar light in the NIR better traces the dominant stellar population by mass
and is less affected by dust extinction.
For distant galaxies, 
either the Balmer break is shifted to the NIR bands 
(ellipticals at $z > 1.5$)
or the optical flux may be obscured by dust (dusty starburst galaxies or AGNs).
These effects produced red optical-to-NIR colors,
and at their most severe extremes result in various red galaxy populations
(e.g., extremely red objects, \citealp{err1988}; distant red galaxies, 
\citealp{franx2003}; 
see also, \citealp{yan2004}, \citealp{wang2012a}, and \citealp{guo2012}).
The most distant Lyman Break Galaxies (LBGs) known ($z > 7$)
cannot be detected at all in most optical bands
because of absorption by the inter-galactic medium, 
and can only be detected in the NIR.
Apart from studies of galaxies,
NIR observations also are important for studies of Galactic dwarf stars
and starforming regions (both are very red 
because of either low surface temperature or dust extinction).
Hence NIR imaging surveys are very valuable
for studying Galactic objects, stellar masses and the distribution of the
dominant stellar population by mass in nearby galaxies,
and for identifying as well as studying the star-forming properties
and the evolved population of distant galaxies.

The Multiwavelength Survey by Yale-Chile
\citep[MUSYC,][]{gawiser2006,taylor2009,cardamone2010}
covers the $30'\times30'$ Extended Chandra Deep Field-South (ECDFS)
with observations in optical and NIR.
The depths of the NIR data from the MUSYC observations, however, 
are too shallow to study the most distant galaxies,
(23.0 mag and 22.3 mag for $J$ and $K$, respectively, 
at $5\sigma$ for point sources) 
and are not comparable to those of the $Spitzer$ IRAC observations
performed by the $Spitzer$ IRAC/MUSYC Public Legacy in the ECDFS survey
\citep[SIMPLE,][]{damen2011}.
According to the current studies of the luminosity function at $z\sim7$
(e.g.,\citealp{ouchi2009}, \citealp{yan2011}),
an L$^*$ galaxy at $z = 7$ would have apparent AB magnitudes of $26--27$ 
in $J$ and $K_S$,
and about one galaxy with $J\sim25$ at $z\sim7$ 
is expected to be found in a field size
similar to that of the ECDFS.
To find LBGs at $z > 7$ for constraining
the very bright end of the $z > 7$ luminosity function \citep{hsieh2012}
and to study the properties of dusty galaxies at $z = 2$--5, 
we initiated the Taiwan ECDFS Near-Infrared Survey (TENIS) in 2007.
This survey comprises extremely deep $J$ and $K_S$ imaging observations
in the ECDFS using the Wide-field InfraRed Camera 
\citep[WIRCam,][]{puget2004} on the Canada-France-Hawaii Telescope (CFHT).
The $5\sigma$ limiting magnitudes for point sources achieve
25.6 and 25.0 mag in $J$ and $K_S$, respectively,
which shows that the TENIS data are by far the deepest NIR data in the ECDFS.

The SIMPLE project provides deep IRAC data 
at the wavelengths of 3.6, 4.5, 5.8, and 8.0$\mu{m}$ in the ECDFS.
In addition, in the central $10'\times15'$ area of the ECDFS, 
the Great Observatories Origins Deep Survey-South 
(GOODS-S, \citealp{giavalisco2004}) project
provides ultra-deep IRAC data  (M.\ Dickinson et al., 2012, in preparation).
These IRAC data are very important for various researches 
especially for high-$z$ studies.
There are many continuum features of distant galaxies 
shifted to wavelengths beyond the $K_S$ band.
For example, 
for galaxies at $z > 0.4$, the rest-frame $1.6\mu{m}$ bump caused by 
the H$^-$ opacity minimum in the stellar photosphere
\citep{se1999,sawicki2002}
and the Balmer break for galaxies at $z > 4.5$
are redshifted out of the $K_S$ band.
Properly detecting these features in IRAC bands can improve
the quality of the photometric redshift estimation.
In addition, the Balmer break is also an important tracer 
of age and stellar mass of high-redshift galaxies.
Moreover, \citet{hsieh2012} show that IRAC data are essential 
for minimizing contamination from Galactic cool stars 
when searching for $z > 7$ LBGs,
and \citet{wang2012a} show that $K$ - IRAC color
is able to pick up the most extremely dust-hidden galaxies
at redshifts between 1.5 and 5.
These studies show the importance of IRAC data for high-$z$ studies.
Measurements of IRAC fluxes for individual sources, however, 
can be easily contaminated by blended neighbors because of the
large IRAC point-spread functions (PSFs).
This issue becomes very serious in deep IRAC surveys
(e.g., the GOODS IRAC survey)
where the surface density of faint objects becomes very high.
To improve the ECDFS IRAC flux measurements,
we have developed a novel deconvolution technique to 
reduce the effect of object blending in the SIMPLE images.
We then combined our $J$, $K_S$ data with IRAC photometry to form
a multiwavelength catalog.

\begin{figure*}
\epsscale{0.9}
\plotone{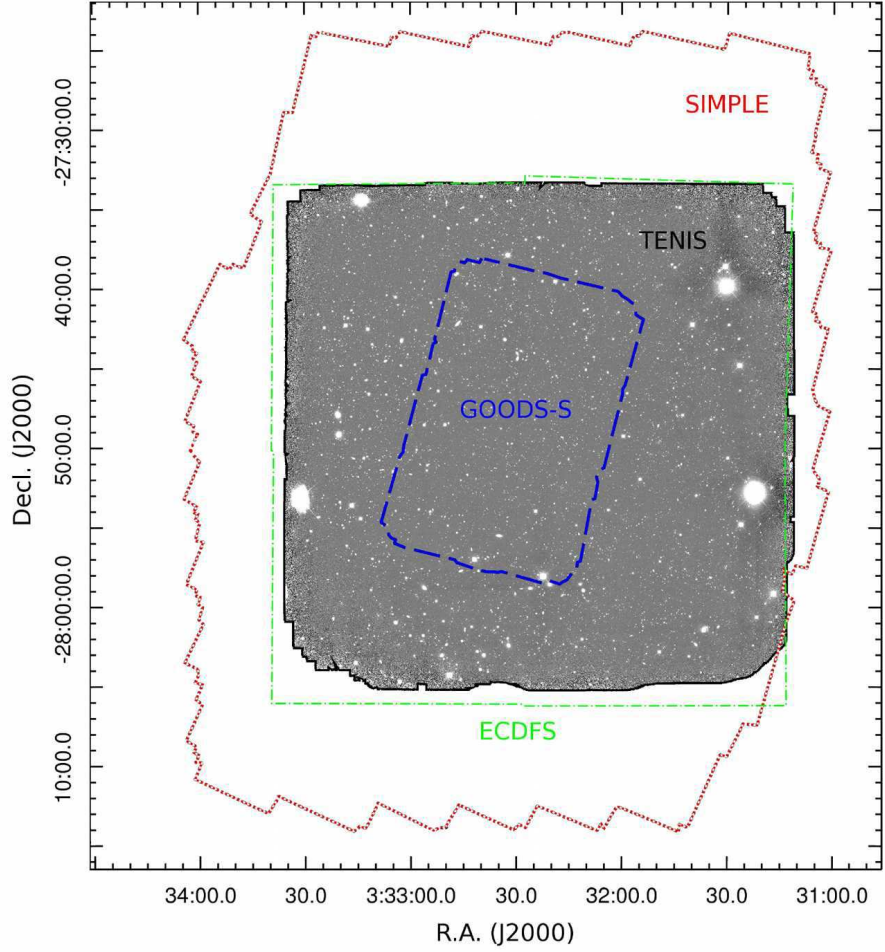}
\caption{The TENIS $J$+$K_S$ image is in grayscale.
The green dot-dashed line indicates the ECDFS field,
the red dotted line indicates the field of the SIMPLE data,
and the blue dashed line indicates the field of the GOODS-S IRAC data.
\label{fov} }
\end{figure*}

We present the survey fields of the TENIS, SIMPLE, 
and GOODS-S IRAC projects in Figure~\ref{fov}. 
We describe the observations of TENIS in Section~\ref{irobs}, 
the procedure of the TENIS data reduction in Section~\ref{datareduction}, 
and the evaluation of the reduction quality in Section~\ref{irquality}.
We describe the photometric catalog of TENIS $J$ and $K_S$ data 
in Section~\ref{jkcat} 
and compare it with other NIR datasets in Section~\ref{isaaccomp}.
In Section~\ref{iracdata}, 
we present our newly developed deconvolution technique
for estimating IRAC fluxes in the ECDFS and examine the performance.
The properties of the combined TENIS and IRAC catalog are discussed in
Section~\ref{mastercatalog}.
In Section~\ref{summary}, we summarize our results.
The images and catalog from this work are available at
the official TENIS website:
\url{http://www.asiaa.sinica.edu.tw/~bchsieh/TENIS/}.
All fluxes in this paper are $f_\nu$.
All magnitudes are in AB system, unless noted otherwise,
where an AB magnitude is defined as AB = 23.9 -- 2.5log($\mu$Jy).

\section{CFHT WIRCam Imaging Observations}\label{irobs}
The TENIS data were taken using WIRCam on the CFHT. WIRCam consists of four 
$2048\times2048$ HAWAII2-RG detectors covering a field-of-view (FOV) of 
$20\arcmin \times 20\arcmin$ with a $0\farcs3$ pixel scale. 
The gap between the four detectors has a width of $45\arcsec$. 
The exposures were dithered to recover the detector gap as well as bad pixels.
We adopted the standard dither pattern provided by the CFHT, 
which distributes the exposures along a ring 
with a default radius of $1\arcmin5$.  
However, we changed the radius every half semester or so, 
to $0.5\times$ to $1.5\times$ of its default value.
This further randomizes the dither footprints and 
minimizes artifacts caused by flat fielding, sky subtraction, 
or crosstalk removal.  
At each dither point, 
we typically co-added two 50-second exposures 
and four 20-second exposures at $J$ and $K_S$, respectively.  
The typical number of dither points were 9 for $J$ and 11 for $K_S$, 
in a dither sequence.  
Therefore, after including the readout time (10 seconds), 
a dither sequence took around 20 minutes to complete.
In our experience, the variation in sky color within such a short period 
is usually sufficiently small to allow for good flat fielding 
and sky subtraction.

The ECDFS has a half-degree size, 
which is larger than the $\sim23\arcmin$--25$\arcmin$ dithered WIRCam FOV.  
To cover the ECDFS, we thus further offset the pointings 
between each dither sequence by $\pm2\farcm5$ along RA and Dec.
In the $J$-band imaging, we offset primarily along the NE--SW direction, 
leaving shallower NW and SE corners.  
In the $K_S$-band imaging, 
we were able to cover all the four corners of the ECDFS. 

\begin{figure*}
\epsscale{1.0}
\plotone{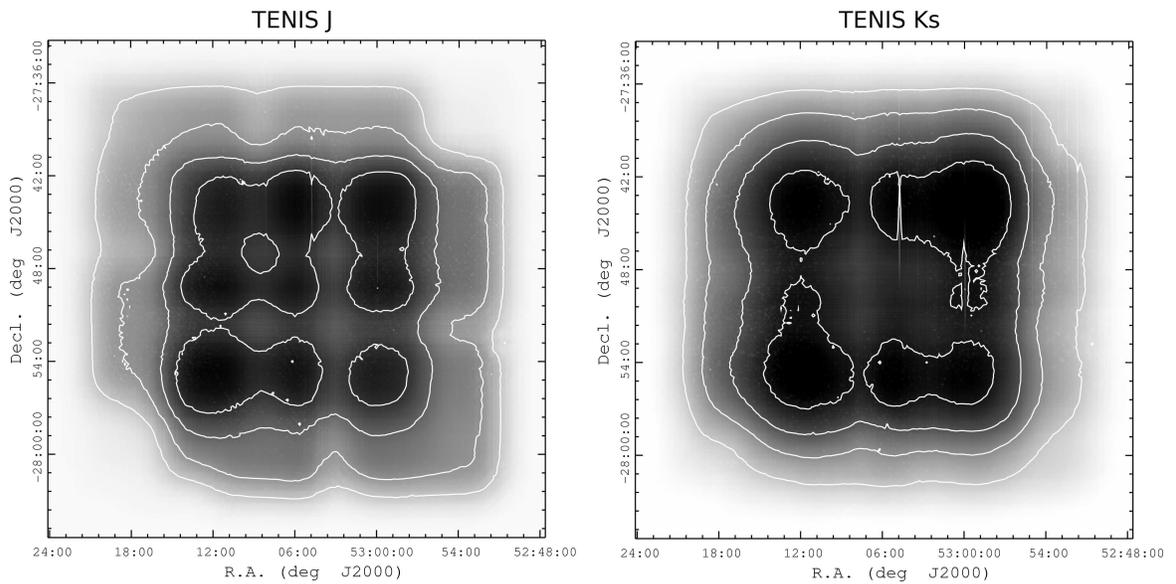}
\caption{Integration maps for the TENIS data.
The left panel is for $J$ and the right panel is for $K_S$.
The grayscale shows the effective integration time in each panel
where darker indicates a longer integration time.
The white contours indicate 20\%, 40\%, 60\%, and 80\%
of the maximum effective integration time,
which is 42.6 hours for $J$ and 39.0 hours for $K_S$.
\label{timemap} }
\end{figure*}

\begin{figure}
\epsscale{1.0}
\plotone{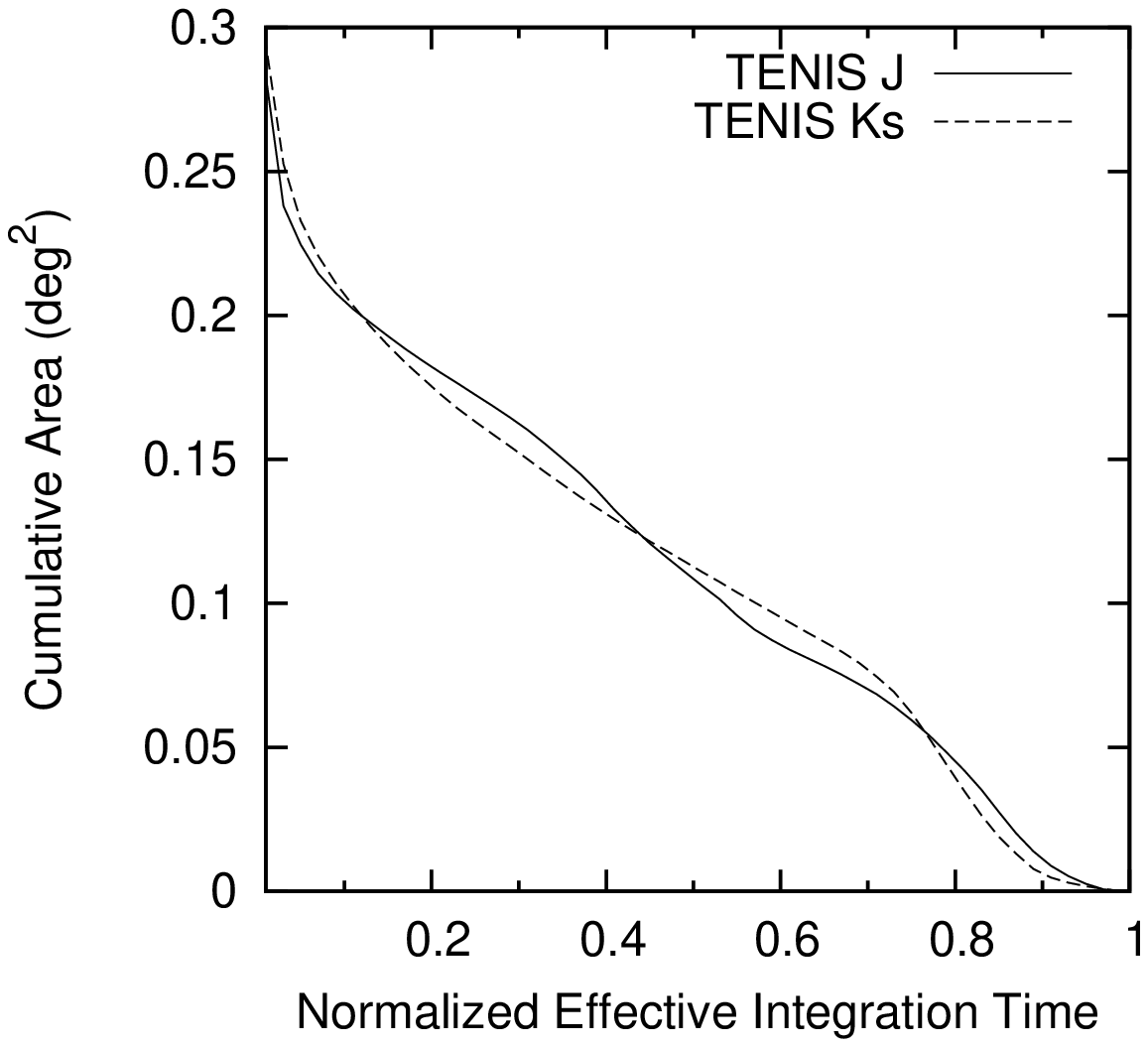}
\caption{Cumulative area in the TENIS $J$ and $K_S$ images
vs. effective integration time.
The maximum effective integration time 
(42.6 and 39.0 hours for $J$ and $K_S$, respectively)
is normalized to be 1.0 for each band.
\label{timevsarea} }
\end{figure}

The observations were carried out by the CFHT in queue mode 
with weather monitoring. 
Most of the data were taken under photometric conditions with similar seeing. 
The average seeing for the observations are $0\farcs8$ and $0\farcs7$ (FWHM) 
for $J$ and $K_S$, respectively. 
For the $J$-band data, we obtained 42.6 hours of on-source integration in 
semesters 2007B and 2008B. 
For the $K_S$-band data, we obtained 25.3 hours of integration 
in semesters 2009B and 2010B. 
In addition, a Hawaiian group led by Lennox Cowie 
obtained 13.7 hours of $K_S$ integration in semester 2010B. 
We include all the 39.0 hours of $K_S$ data here. 
The integration maps for the $J$ and $K_S$ data 
are shown in Figure~\ref{timemap},
and the cumulative areas in the TENIS $J$ and $K_S$ images
as functions of effective integration time
are shown in Figure~\ref{timevsarea}.

\section{Data Reduction}\label{datareduction}
The TENIS $J$ and $K_S$ data were processed 
using an Interactive Data Language (IDL) based reduction pipeline.
The details of the pipeline are described in \citet{wang2010}, 
who also dealt with extremely deep WIRCam imaging.
\footnote
{Also see \url{http://www.asiaa.sinica.edu.tw/~whwang/idl/SIMPLE.}}
The reduction here is nearly identical to that in \citet{wang2010}.

We first grouped the images and only processed images 
from the same dither sequence and the same detector at a time.  
The dithered raw images were first median-combined
to produce an initial flat field.  
Then objects were detected on each flattened image
and were masked on the associated raw image.  
For an image, a better flat-field image was then created 
by median-combining the rests of the object-masked raw images 
in the dither sequence.  
This way, every image has an associated flat field, created by avoiding itself. 
This is the major difference between the reduction here and
that in the early data release of \citet{wang2010}.  
In the earlier versions of the \citet{wang2010} pipeline, 
all images in a dither sequence were median combined 
to form a flat field image, after detected objects were masked.  
This produces an extremely small 
but statistically significant overestimate of the flat field 
at the locations of faint objects that are undetected in single exposures. 
Because the sky background is very bright, 
this small error in flat field translates to a larger error 
in the final fluxes of faint objects. 
This is discussed in details in the later data release of Wang et al. 
(\url{http://www.asiaa.sinica.edu.tw/~whwang/goodsn\_ks}). 
We would like to draw to the community's attention
that a similar error may exist in other datasets 
if a similar flat-fielding method is adopted.  
Our reduction here avoids the above problem.

After the above flat-fielding, usually the images are sufficiently flat 
and only a constant background subtraction is needed.
Occasional residual sky structure caused 
by rapidly changing sky color was further subtracted 
by fitting fifth-degree polynomial surfaces to the image background. 

On each WIRCam detector, there is crosstalk among the 32 readout
channels ($2048\times64$ pixels each).  
The crosstalk has different strength within the entire detector (32 channels), 
and within each of the four video boards (eight channels each). 
For every flattened and background subtracted image, 
we removed the 32-channel crosstalk by subtracting the median combination 
of the 32 $2048\times64$ object-masked stripes. 
A similar procedure is then repeated to remove the 8-channel crosstalk.
This greatly suppresses the crosstalk in the images, 
and only very weak residual effect can be seen 
in the final ultradeep stack around several tens of the brightest objects 
in the ECDFS.

To correctly stack the dithered images, 
optical distortion needs to be corrected.
Our reduction pipeline followed the method developed by \citet{anderson2003}
to derive the distortion function.  
The pipeline first detected objects in all dithered images, 
and then calculated the spatial displacement caused by the dithering 
for each object.  
Such displacement as a function of position in the images is actually
the first-order derivative of the optical distortion function.  
The pipeline approximated the displacement function 
with polynomial functions of X and Y, 
and integrated them back to obtain the distortion function.  
We referred to Wang et al. (2010) for more detailed discussion 
about this technique.  

To obtain absolute astrometry and to project the images onto the sky, 
we compared the distortion-corrected object positions 
with the source catalog produced 
by the \emph{HST} Galaxy Evolution from Morphology and SEDs 
\citep[GEMS,][]{rix2004,caldwell2008} survey.  
By forcing the object positions to match those in the GEMS catalog, 
we computed a single two-dimensional, 
third-degree transformation function 
that contains all the effects including the distortion, 
sky projection, and absolution astrometry, for each dithered image.  
Therefore, in the entire reduction, 
each image only underwent one geometric transform. 
This minimizes the impact of image smearing caused by the transform.
The transformed images can then be stacked to form a deep, 
astrometrically correct image. 

Before images were stacked, photometry is carried out on individual exposures,
and compared among the exposures.  
This allows to adjust relative zero-points of the individual exposures.  
After the images in a dither sequence were stacked, 
absolute zero-point calibration was made by matching the fluxes 
in apertures of $5\arcsec$ in diameter with the ``default magnitudes''
in the point source catalog of the Two Micron All Sky Survey 6
\citep[2MASS,][]{skrutskie2006}. 
\footnote{2MASS is a joint project of the University of Massachusetts and the 
Infrared Processing and Analysis Center/California Institute of Technology, 
funded by the National Aeronautics and Space Administration and the 
National Science Foundation.}
Since the magnitudes in the 2MASS catalog are in the Vega system,
we converted the 2MASS magnitudes to fluxes based on the Vega zero-magnitudes 
in $J$ and $K_S$ of 1594 Jy and 666.8 Jy, respectively,
as provided by 2MASS \citep{cohen2003}.
All the stacked, calibrated images from single-dither sequences 
and from the four WIRCam detectors were then mosaicked for form a wide-field, 
ultradeep image.

\section{Reduction Quality}\label{irquality}
\subsection{Astrometry}\label{astroqua}
\begin{figure}
\epsscale{1.0}
\plotone{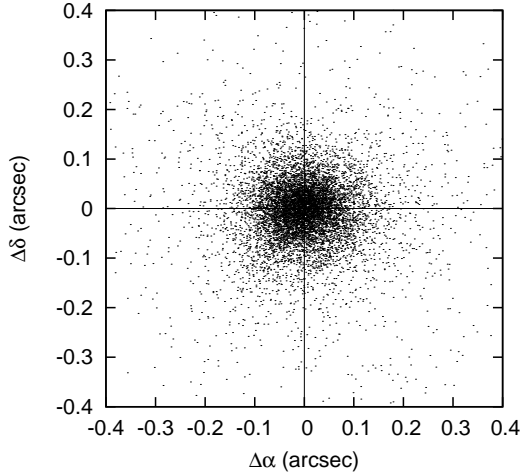}
\caption{Relative astrometric offsets between
the TENIS $J$+$K_S$ image and the GEMS ACS catalog.
We selected over 9,000 compact sources with good detections
(i.e., S/N $> 20$ and FWHM $< 1\farcs2$) for the comparison.
The systematic offsets in R.A. and Decl. are $0\farcs0016$ and $0\farcs0035$,
which are negligible as compared to the rms offsets,
which are $0\farcs087$ and $0\farcs080$ for R.A. and Decl., respectively.
\label{astrometry1} }
\end{figure}

\begin{figure*}
\epsscale{1.0}
\plotone{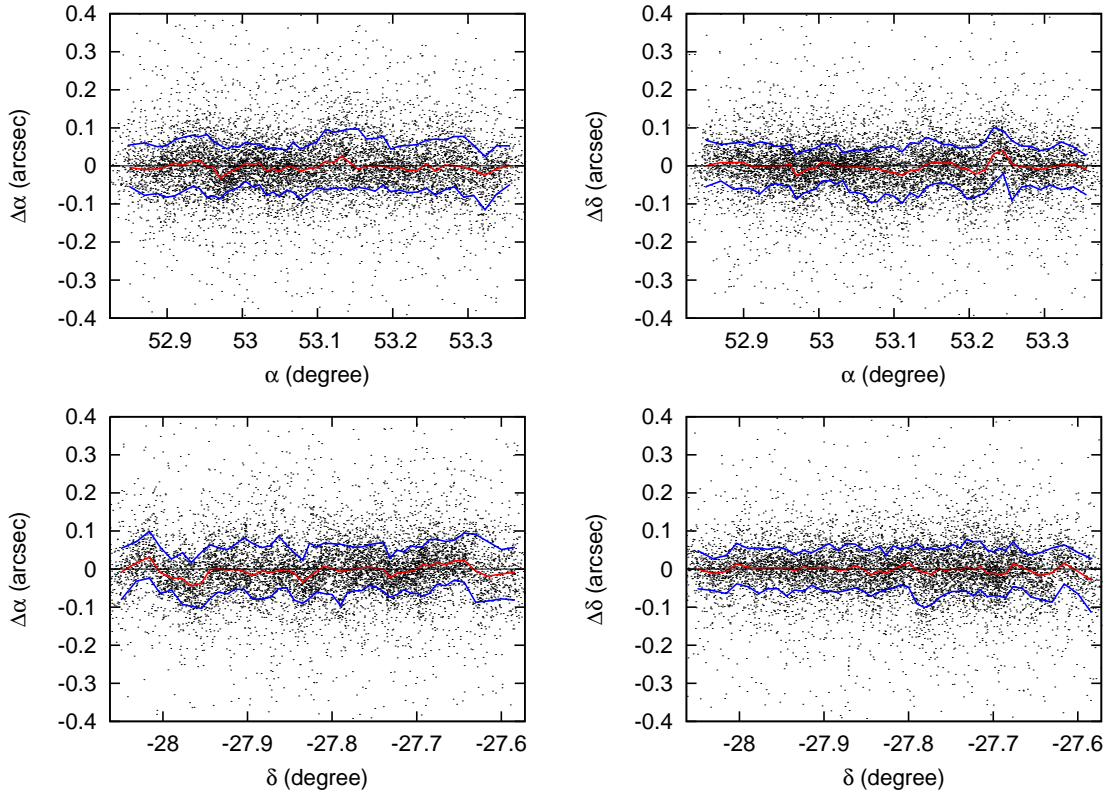}
\caption{Same as Figure~\ref{astrometry1}
but as function of R.A. and Decl.
The red line indicates the running median
and the blue lines indicate the upper and lower 68th percentiles.
\label{astrometry2} }
\end{figure*}

The GEMS survey used the Advanced Camera for Surveys (ACS) on the 
Hubble Space Telescope (\emph{HST}) to image nearly the entire ECDFS.
As described in the previous section, 
we forced our astrometry to match that of the GEMS ACS catalog.  
Here we compare how well our astrometry matches that of the GEMS catalog.
More than 9,000 compact sources with good detections 
(S/N $> 20$ and FWHM $< 1\farcs2$)
are selected for this comparison.
Figures~\ref{astrometry1} and ~\ref{astrometry2} show
the relative astrometric offsets between the TENIS $J$+$K_S$ image
and the GEMS ACS catalog.
As shown in Figure~\ref{astrometry1},
the systematic offsets in R.A. and Decl. 
are $0\farcs016$ and $0\farcs0035$, respectively. 
These offsets are negligible compared with the rms scatter,
which are $0\farcs087$ and $0\farcs080$ for R.A. and Decl., respectively.
On the other hand, there are position-dependent systematic offsets.
In the bottom-left panel of Figure~\ref{astrometry2},
there is a wiggling structure with a scale of $\sim0.05$ degree,
which is comparable to the FOV of the ACS Wide-field Camera
\citep[WFC,][FOV = $202''\times202''$]{ford2003}.
Similar but less obvious structures can also be found 
in the other panels of Figure~\ref{astrometry2}.
It is highly likely that these position-dependent systematic offsets 
are internal to the GEMS catalog.

According to \citet{caldwell2008}, 
the astrometry of GEMS catalog is registered to that of
the Classifying Objects through Medium-Band Observations,
a spectrophotometric 17-filter survey \citep[COMBO-17,][]{wolf2001}. 
The absolute astrometric accuracy of the GEMS catalog is therefore limited by 
the astrometric quality of the COMBO-17 catalog, 
which is better than $0\farcs15$ 
but which may be greater than $0\farcs3$ in some localized regions.
As we calibrated our astrometry using the GEMS catalog,
the absolute astrometric accuracy of the TENIS data is therefore
also limited by that of the COMBO-17 data.
We could match the TENIS astrometry directly to the COMBO-17 data.
However, the major scientific goal of the TENIS project
is to find LBGs at $z>7$, which needs the ultra deep GEMS data.
We therefore matched the TENIS astrometry to that of the GEMS data 
rather than that of the COMBO-17 data
because it makes combining the TENIS data and the GEMS data much easier.

\subsection{Photometry}\label{photqua}
\begin{figure}
\epsscale{1.0}
\plotone{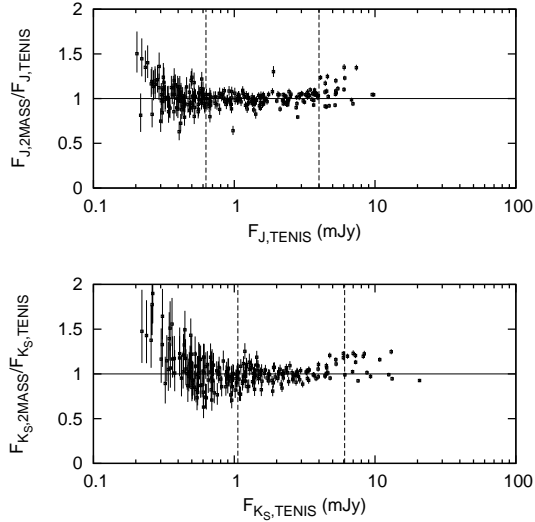}
\caption{Flux ratios between 2MASS and TENIS in $J$ and $K_S$.
The two vertical dashed lines in each panel indicate
the flux range for the flux calibration.
We note that the outliers between the two vertical dashed lines
are excluded in the flux calibration.
The flux errors of the TENIS data are also shown horizontally
but they are too small to be visible.
Hence they have negligible contribution to the vertical error bars.
The flux calibration quality of the TENIS data is therefore
completely dominated by the 2MASS flux errors.
\label{tenis-2mass} }
\end{figure}

As mentioned in Section~\ref{datareduction},
the TENIS WIRCam $J$ and $K_S$ fluxes were calibrated
using $5''$ diameter apertures and matched
to the fluxes in the 2MASS point source catalog.
In Figure~\ref{tenis-2mass}, 
we show the flux ratios between 2MASS and TENIS in $J$ and $K_S$,
where the flux range between the two vertical dashed lines in each panel
indicates that used for the calibration.
Fluxes brighter than this range suffer from nonlinearity issues in WIRCam,
whereas fluxes fainter than this range are subjected 
to selection effects due to the much shallower detection limits of 2MASS.
The error-weighted means of the objects in the chosen calibration flux ranges
are 1.0002$\pm$0.003 and 1.0006$\pm$0.005
for $J$ and $K_S$, respectively,
resulting in an overall flux calibration quality 
that is good to 0.3\% and 0.5\% for $J$ and $K_S$, respectively.
We note that the flux errors of the TENIS data are also shown horizontally 
in Figure~\ref{tenis-2mass} 
but are too small to be visible,
and hence they make negligible contributions to 
the photometric error budget compared with the vertical error bars.
The flux calibration quality of the TENIS data
is therefore completely dominated by the 2MASS flux errors.

We also investigated the photometric uniformity in our TENIS images
by subdividing the TENIS $J$ and $K_S$ images into four quadrants
and then comparing the TENIS fluxes with the 2MASS fluxes
as we did for the entire images.
For quadrants 1 through 4 in $J$, the error-weighted flux ratios are
0.979 $\pm$ 0.007 (21 sources), 0.997 $\pm$ 0.006 (25 sources),
1.014 $\pm$ 0.005 (43 sources), and 1.005 $\pm$ 0.008 (22 sources), 
respectively.
For those in $K_S$, they are 0.998 $\pm$ 0.009 (19 sources),
1.033 $\pm$ 0.007 (27 sources), 1.003 $\pm$ 0.009 (27 sources),
0.978 $\pm$ 0.011 (11 sources), respectively.
The standard deviations of the four measured offsets are 1.3\% and 2.0\%
for $J$ and $K_S$, respectively.
This suggests that the photometric gradients over size scales of $\sim15'$
are less than 0.013 and 0.02 mag in the TENIS $J$ and $K_S$ images,
respectively;
these measured 1.3\% and 2.0\% are fairly good
as compared to other NIR extragalactic deep surveys
over similar size scales
(e.g., 2\% in the COSMOS survey; \citealp{capak2007}).

\section{The $J$ and $K_S$ photometric catalog}\label{jkcat}
To produce a source catalog that is as complete as possible,
we decide to perform object detection in a high S/N image generated by
combining the TENIS $J$ and $K_S$ images.
Such a combination has to take into account
the different integration time distributions in $J$ and $K_S$.
A straightforward method is to weight the pixels
by their associated integration times and then combine the images.
However, our $J$ integration time is approximately $1.32\times$
longer than the $K_S$ integration time,
but objects are generally brighter (in our map unit, which is $\mu$Jy) at $K_S$.
Hence, a direct integration time weighted combination is less optimal
in terms of combined S/N for most objects.
We therefore normalized the $J$ and $K_S$ integration times
by artificially reducing the $J$ integration time by a factor of 1.32
(i.e., downweighting the $J$ image),
and then performed the integration time weighted $J$+$K_S$ combination.

We used SExtractor version 2.5.0 \citep{ba1996}
to detect objects and measure their fluxes in the TENIS WIRCam images.
The ``FLUX\_AUTO'' values of the SExtractor output are chosen
for the flux measurements.
The double-image mode of SExtractor was performed to 
detect objects in the $J$+$K_S$ image
while measuring fluxes in the original $J$ and $K_S$ images 
at the locations of the $J$+$K_S$ objects. 
The most important SExtractor parameters 
we used are listed in Table~\ref{sexpar}.

\begin{deluxetable}{lc}
\tabletypesize{\scriptsize}
\tablecolumns{2}
\tablewidth{0pt}
\tablecaption{SExtractor Parameters
\label{sexpar}}
\tablehead{
\colhead{Parameter} & \colhead{Value}}
\startdata
DETECT\_MINAREA & 2 \\
DETECT\_THRESH & 1.3 \\
ANALYSIS\_THRESH & 1.3 \\
FILTER & Y \\
FILTER\_NAME & gauss\_2.0\_3x3.conv \\
DEBLEND\_NTHRESH & 64 \\
DEBLEND\_MINCONT & 0.00001 \\
CLEAN & Y \\
CLEAN\_PARAM & 0.1 \\
BACK\_SIZE & 24 \\
BACK\_FILTERSIZE & 3 \\
BACK\_TYPE & AUTO \\
BACKPHOTO\_TYPE & LOCAL \\
BACKPHOTO\_THICK & 40 \\
WEIGHT\_TYPE & MAP\_WEIGHT \\
WEIGHT\_THRESH & 20 \\
\enddata
\end{deluxetable}

The flux errors provided by SExtractor are derived 
from the background noise directly,
and the noise covariance between pixels 
(i.e., correlated error produced by image resampling, 
and to a lesser degree, faint undetected objects) is ignored.
To mitigate against the latter effect, 
we calibrated the errors using the following procedure.
First, the fluxes and the flux errors were re-measured
using SExtractor with $2''$ diameter apertures.
We then convolved the source-masked image 
with a $2''$ diameter circular top-hat kernel,
and calculated the rms in a $10''\times10''$ area 
around each pixel on the convolved image.
The ratio between the aperture photometric flux error provided by SExtractor
for a certain object and the rms value
around the same position in the convolved image
is the correction factor for its flux error.
The median value of the correction factors was computed to be
the general correction factor for all the sources.
For the TENIS $J$- and $K_S$-band data, 
the general correction factors are very similar, which is 1.27.
The same procedure was repeated with different aperture sizes.
We found that the correction factors are very stable
over different aperture sizes,
which is consistent with the experience in \citet{wang2010}.
Therefore we applied this factor to the ``FLUXERR\_AUTO'' values
for all the objects to account 
for the effects of noise correlation and confusion.
We note that this factor does not apply to the flux measurements.

We then estimated the aperture correction for the AUTO aperture.
Since we calibrated the WIRCam data to 2MASS using a $5''$ diameter aperture,
we just need to derive the correction factor 
between the AUTO aperture and the $5''$ aperture.
After comparing the fluxes derived using these two different apertures,
we found that they are in very good agreement with each other for most objects.
Because the photometry derived using AUTO apertures have better S/N
for objects with various different morphologies and fluxes 
as compared to the $5''$ fluxes,
we decided to use the original ``FLUX\_AUTO'' values 
as the final flux measurements in our catalog.

\section{Comparison with the GOODS-S/ISAAC Data}\label{isaaccomp}
\begin{figure*}
\epsscale{1.0}
\plotone{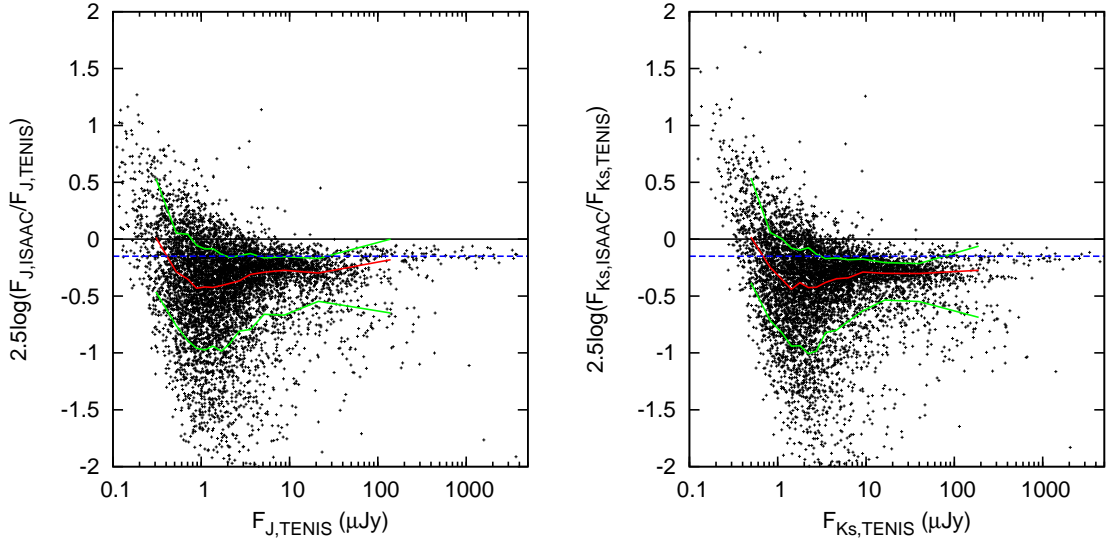}
\caption{Photometry comparisons between the TENIS and ISAAC catalogs.
Flux ratios between the two catalogs versus the TENIS fluxes are plotted.
The left panel is for $J$ and the right panel is for $K_S$.
The red line indicates the running median
and the green lines indicate the upper and lower 68th percentiles.
The blue dashed line indicates a -0.15 mag difference.
The results show that the ISAAC fluxes for bright stars
are about 10\% to 15\% less than the TENIS fluxes.
\label{tenis-isaac_offset} }
\end{figure*}

\begin{figure*}
\epsscale{1.0}
\plotone{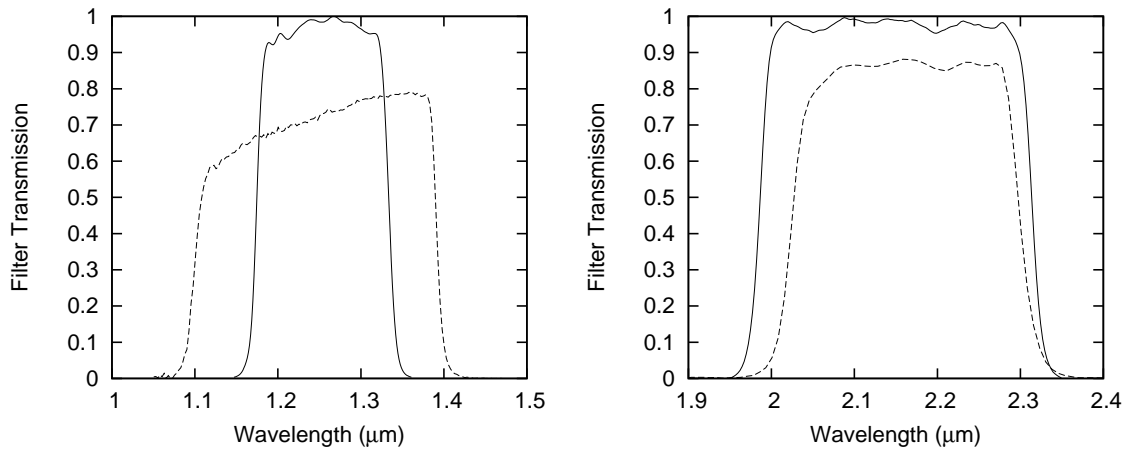}
\caption{Transmission curves for the WIRCam and ISAAC filters.
The left panel is for $J$ and the right panel is for $K_S$.
The solid curve is for WIRCam and the dashed curve is for ISAAC
in each panel.
\label{transcurve} }
\end{figure*}

There are deep near-infrared imaging observations of the GOODS-S region
using the Infrared Spectrometer And Array Camera (ISAAC) 
on the Very Large Telescope (VLT) in $J$, $H$, and $K_S$.
The $K_S$-selected catalog is published by \citet{retzlaff2010}.
According to the sensitivities provided by \citet{retzlaff2010},
the depths of the ISAAC data are comparable with that of TENIS.
We have checked the photometric consistency 
and the data quality difference between the two catalogs.
In Figure~\ref{tenis-isaac_offset},
we compare the TENIS $J$ and $K_S$ photometry 
with the total magnitudes in the ISAAC catalog.
The ISAAC fluxes for bright stars are approximately 10\% and 15\%
less than the TENIS fluxes at $J$ and $K_S$, respectively.
We checked whether these large differences could be caused by the 
differences in the filter systems used, as shown in Figure~\ref{transcurve}.
We found that the different passbands of the two filter systems 
can produce only $\sim3\%$ differences in fluxes
by examining the color of stars 
from Kurucz models \citep[ATLAS9;][]{kurucz1993}.
It is also worth noting that 
both catalogs are not corrected for Galactic extinction,
although the correction values, only 0.008 mag ($J$) and 0.003 mag ($K_S$),
are too small to explain the $>10\%$ flux differences.

\begin{figure*}
\epsscale{1.0}
\plotone{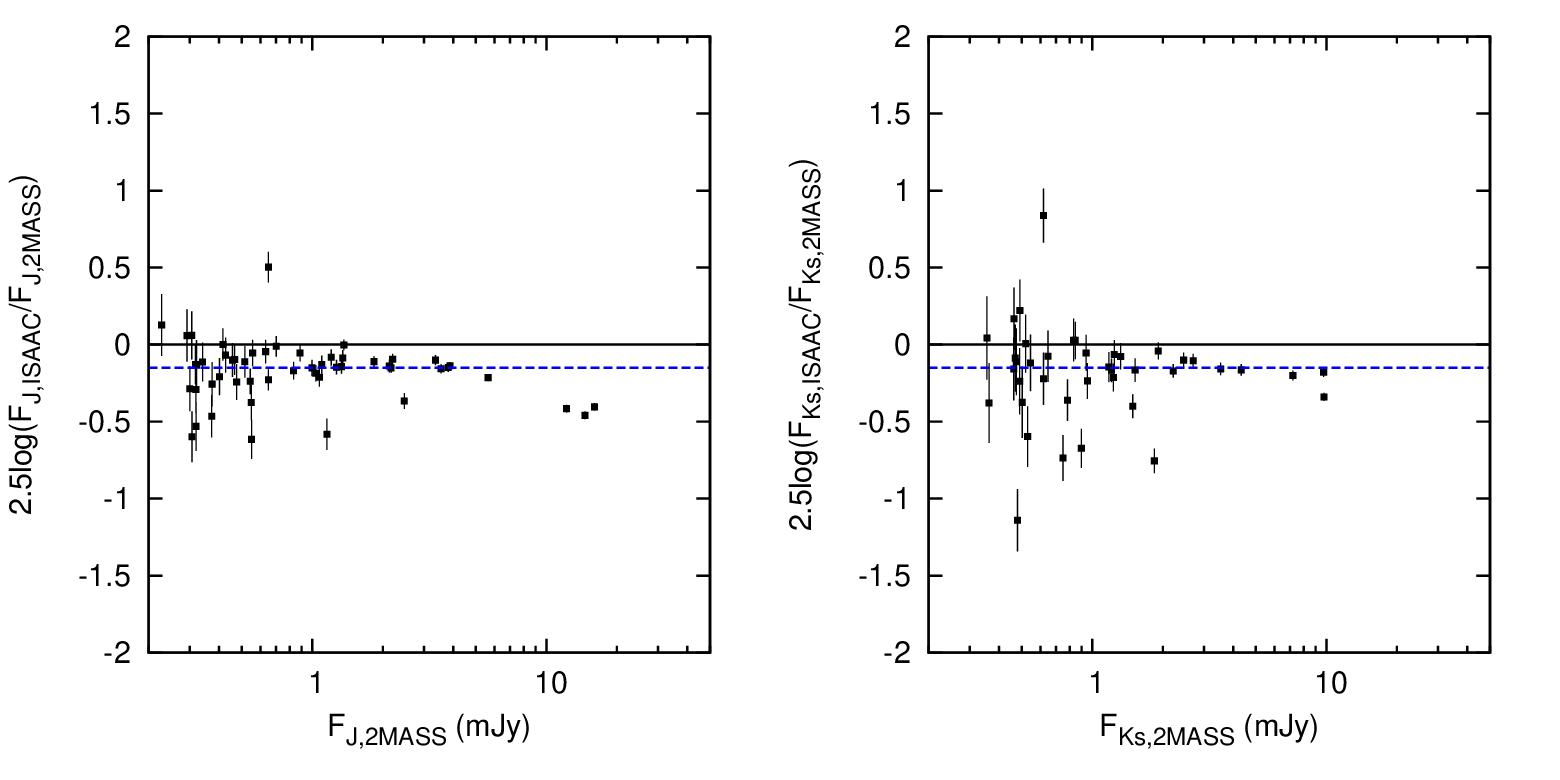}
\caption{Same as Figure~\ref{tenis-isaac_offset} but for 2MASS and ISAAC.
The blue dashed line indicates a -0.15 mag difference.
The flux differences are consistent with
that shown in Figure~\ref{tenis-isaac_offset},
which suggests that the flux discrepancies 
between the TENIS and ISAAC catalog are due to
different zeropoint calibration methods.
\label{isaac-2mass} }
\end{figure*}

We calibrated the TENIS photometry using the 2MASS catalog.
The ISAAC observations, however, were calibrated with standard stars.
We therefore directly compared the ISAAC catalog with the 2MASS catalog
to see if the flux differences are due to 
different zeropoint calibration methods used in the two catalogs.
The result is shown in Figure~\ref{isaac-2mass}.
The ISAAC total fluxes are $\sim10$\% to 15\%
lower than the 2MASS default fluxes, 
consistent with the differences between the TENIS and ISAAC fluxes.
In addition, \citet{retzlaff2010} mentioned that
a significant bias ($\leq$0.1mag) is visible 
when comparing ISAAC to the 2MASS catalog.
Hence we conclude that the flux discrepancies 
between the TENIS and ISAAC catalogs
are due to different zeropoint calibration methods.

\begin{figure*}
\epsscale{1.0}
\plotone{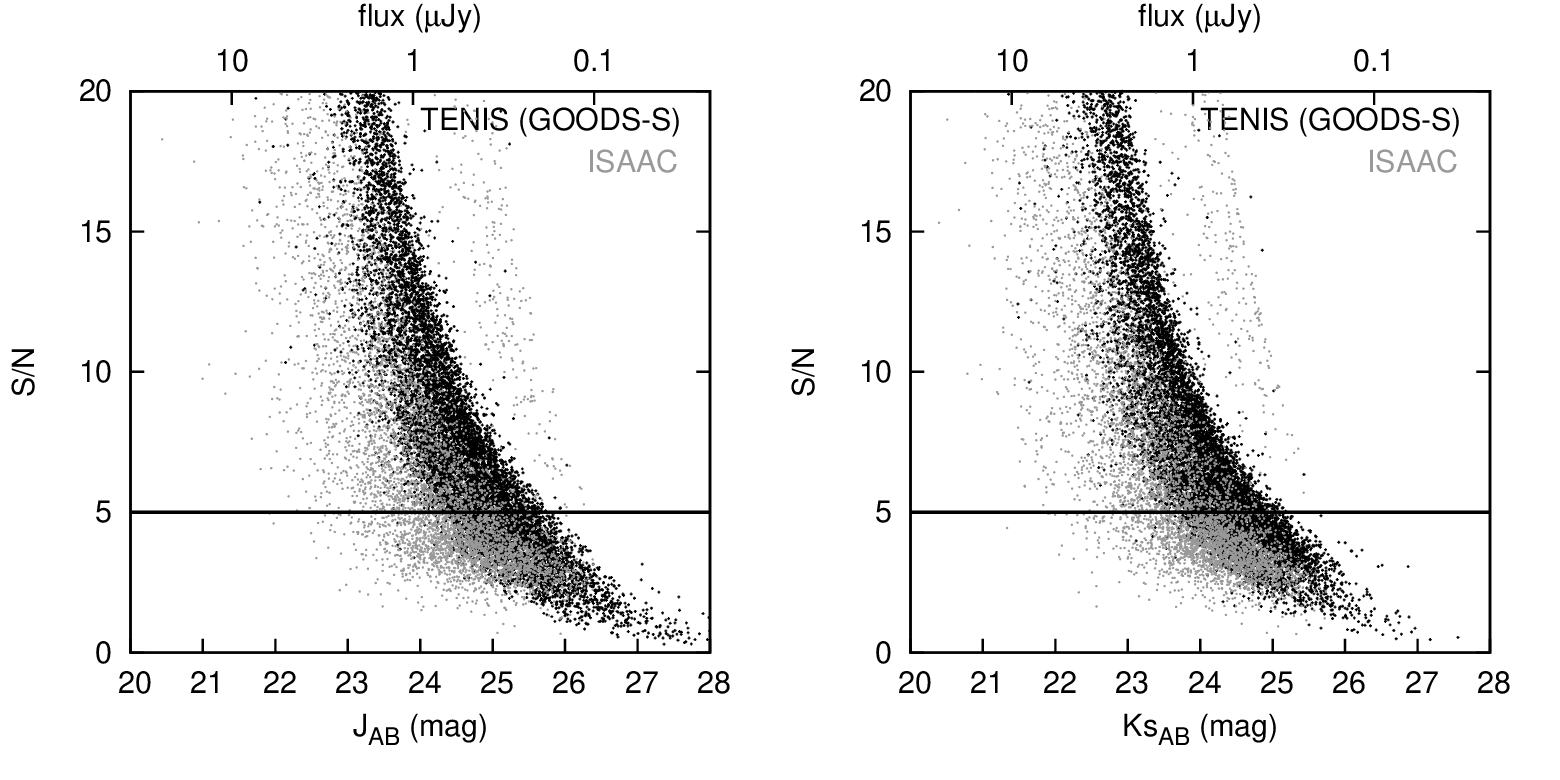}
\caption{S/N vs. Magnitude plots for the TENIS and ISAAC data.
The left panel is for $J$ and the right panel is for $K_S$.
The dark black dots indicate objects in the GOODS-S region in the TENIS catalog
and the gray dots indicate objects in the ISAAC catalog.
The horizontal line in each panel marks the S/N level of 5.
Based on this comparison, the TENIS data are about 0.5mag deeper than
the ISAAC data in both $J$ and $K_S$.
\label{tenis-isaac_depth} }
\end{figure*}

To compare the depth, we plot the magnitude vs. S/N 
for the TENIS and ISAAC data in Figure~\ref{tenis-isaac_depth}.
The $5\sigma$ limiting magnitudes for point sources of the ISAAC data claimed by
\citet{retzlaff2010} are about 25.0 and 24.4 for $J$ and $K_S$, respectively,
which are consistent with what are shown in Figure~\ref{tenis-isaac_depth}.
According to this figure, the TENIS data are about 0.5 mag deeper than
the ISAAC data in both $J$ and $K_S$.
If the abovementioned zeropoint differences are considered,
the depth differences would be further increased to $>0.6$ mag.
We note that the Cosmic Assembly Near-infrared Deep Extragalactic Legacy Survey 
\citep[CANDELS,][]{grogin2011,koekemoer2011} provides 
a deeper $J$-band ($F125W$) dataset in the GOODS-S region 
(5$\sigma$ limiting magnitudes for point sources is $\sim27$ mag).
We therefore conclude that our TENIS catalog is by far
not only the deepest NIR dataset in ECDFS,
but also the deepest $K_S$ dataset even in the narrow GOODS-S region.

\section{IRAC Photometry}\label{iracdata}
The SIMPLE survey \citep{damen2011} provides deep IRAC observations 
covering the entire ECDFS with the $10\arcmin \times15\arcmin$ 
ultradeep GOODS-S IRAC 
(M.\ Dickinson et al., 2012, in preparation) mosaics in its center.
According to \citet{damen2011},
the $5\sigma$ limiting magnitudes for point sources
typically are 23.8, 23.6, 21.9, and 21.7 
for 3.6$\mu$m, 4.5$\mu$m, 5.8$\mu$m, and 8.0$\mu$m, respectively.
For the ultradeep GOODS-S IRAC region,
the $5\sigma$ limiting magnitudes for point sources
are 26.1, 25.5, 23.5, and 23.4
for 3.6$\mu$m, 4.5$\mu$m, 5.8$\mu$m, and 8.0$\mu$m, respectively,
according to \citet{dahlen2010}.
As mentioned in Section~\ref{introduction},
IRAC data are valuable for studies from Galactic objects 
to the high-redshift universe.
Combining the SIMPLE IRAC and TENIS data can broaden the use
of the TENIS catalog.

\subsection{Basic Principle}\label{principle}

The major difficulty in combining the IRAC and WIRCam data is 
the relatively large IRAC PSFs ($1\farcs5$ to $2\farcs0$) 
compared with of the WIRCam data ($0\farcs8$).
As a consequence, measurements of total fluxes of objects in the IRAC images 
require photometric apertures that are much larger than the WIRCam apertures.  
Given the ultradeep nature of the SIMPLE and GOODS-S IRAC images 
and the high surface densities of objects, 
large apertures mean that the photometric accuracy is highly subject 
to nearby bright objects as well as 
the large number of faint confusing sources. 
Because the blended objects can have very different morphology and color,
accounting for this problem is crucial.

Many intensive efforts have gone into better estimating fluxes in crowded, 
low-resolution images.
The most recent methods rely on utilizing the positional 
and morphological information of objects detected 
in a high resolution image in a different waveband
\citep[e.g.,][]{grazian2006,laidler2007,wang2010,mclure2011}.
By assuming that the intrinsic morphology of an object is identical
in the two wavebands (i.e., no color gradient from center to edge) 
and taking into account differences in the PSFs, 
one can model how an object would look like in the low
resolution image based on its observed morphology in the high resolution image.
Using a minimum $\chi^2$ method, \citet{grazian2006} 
and \citet{laidler2007} were able to model all blended objects simultaneously 
and thus estimate their fluxes in the low resolution image. 
Instead of minimizing $\chi^2$, 
another approach is to minimize residual fluxes between the model 
and the observed low resolution image \citep{wang2010}. 
This is the basic idea of CLEAN deconvolution 
in radio imaging \citep{hogbom1974}.  

In this work, we follow the CLEAN approach of \citet{wang2010}. 
However, we greatly relax the assumption on object morphology. 
In \citet{wang2010}, 
a model of the IRAC image of a galaxy is generated based on the PSFs 
and its high-resolution WIRCam image, 
and the model is then iteratively subtracted (``CLEANed'') 
from the real IRAC image.
Here, we directly CLEAN the IRAC image of a WIRCam detected galaxy 
using the IRAC PSF, and we allow the cleaning position 
to move around its WIRCam position. 
In other words, 
our method is essentially the same as CLEAN deconvolution in radio imaging, 
with nearly no restrictions except for the locations where CLEAN can happen.
We named this method ``IRACLEAN'' 
because it is designed for estimating IRAC fluxes.

\subsection{Methodology}

\subsubsection{PSF Construction}\label{psfgen}
The IRAC PSFs were generated using bright isolated point sources 
in each IRAC channel.
The size of the PSF images is $1'\times1'$ 
in order to cover the outer structure of the extended IRAC PSFs, 
especially for 8.0$\mu$m.
To avoid saturated sources, 
we chose intermediate brightness objects
and checked their images carefully.
We also checked their optical counterparts in the extremely high resolution
\emph{HST} ACS images to make sure that they are really point-like.
The final PSF is constructed by calculating the flux weighted mean of the PSFs
of these objects with a $3\sigma$ clip.
There are, however, still many objects within $1\arcmin \times1\arcmin$ around 
these sources even if they are relatively ``isolated''.
This biases the wing of the constructed PSF.
We therefore ran SExtractor on the IRAC images 
to generate object masks and then masked all the nearby objects
when generating the PSFs.
This masking procedure, however, 
may create many ``holes'' in the constructed PSF image.
To overcome this, 
sometimes more than 20 objects have to be used for constructing one PSF.
The PSFs of stars in the native IRAC images would be under-sampled
because of the relatively large pixel size of IRAC
($1\farcs2$ per pixel for the IRAC PSFs with FWHMs of
$1\farcs66$ to $1\farcs98$).
Without doing a sub-pixel centering, 
the constructed PSF may be artificially broadened.
Since the SIMPLE image has been resampled to match the TENIS image
($\sim 0\farcs3$ per pixel),
simply finding the location of a peak 
is very similar to do a sub-pixel centering, 
and the centering accuracy of the constructed PSF is better than
9\% to 7.6\% ($\pm 0\farcs15$, from 3.6$\mu$m to 8.0$\mu$m) of the IRAC FWHMs.
Hence the broadening effect of the stacked PSF is negligible.
We note that the PSFs in the SIMPLE images are position-dependent,
and the procedure that we used to deal with this issue is described in
Section~\ref{7psf}.

\subsubsection{Object Identification}\label{objident}
We used the TENIS $J$+$K_S$ image as a prior
to estimate the IRAC fluxes around the locations of 
$J$+$K_S$ detected objects.
The location information is contained in the SExtractor
``segmentation map'' of the TENIS $J$+$K_S$ image. The segmentation map 
tells which detected object (or no object) a pixel is associated with.
For a given $J$+$K_S$ detected object, we only clean its IRAC fluxes 
at its associated segmentation map pixels.
Therefore, the boundary of an object defined in the segmentation map
is equivalent to ``CLEAN window'' in radio imaging.
In addition, since we allow any pixels within the boundary
of an object to be CLEANed, 
we do not assume any morphology for that object except
for its outer extent as defined by the $J$+$K_S$ image. 
This has the advantage of 
allowing changes in morphology at different wavelengths.
We should note that changing the SExtractor detection threshold setting
for the $J$+$K_S$ image (i.e., DETECT\_THRESH) would
enlarge and shrink the segmentation area, 
which might affect the IRACLEAN performance.
To ensure IRACLEAN working properly, 
the detection threshold should be set as low as possible.
Lowering the detection threshold down too much, however,
would also increase the spurious rate significantly.
We found that the IRACLEAN fluxes with different DETECT\_THRESH values
agree with each other within 3\% as long as DETECT\_THRESH $\leq$ 1.5,
and the spurious rate increases dramatically with DETECT\_THRESH $\leq$ 1.2
(see Section~\ref{spurious} for details about spurious rate).
Setting DETECT\_THRESH = 1.3 (see Table~\ref{sexpar}) is a good balance
between IRACLEAN performance, source completeness, and spurious rate.

\subsubsection{IRACLEANing Objects}\label{clean}
We first made mosaics of the IRAC images and resampled them to match 
the pixel size of the TENIS $J$+$K_S$ image ($0\farcs3$).
The IRACLEAN process always starts at an IRAC pixel 
with the highest ``absolute'' value (i.e., the value can be negative)
measured within a $9\times9$-pixel box ($F_{AP9}$, hereafter),
and this pixel ($P_{decon}$, hereafter) must have been registered 
to one object in the $J$+$K_S$ segmentation map.
We chose a $9\times9$-pixel ($2\farcs7 \times 2\farcs7$) box 
as the aperture size because it delivers the best S/N.
Since the IRAC image ($0\farcs6$ per pixel) has been resampled 
to match the TENIS $J$+$K_S$ image ($0\farcs3$ per pixel),
simply moving a window (i.e., $F_{AP9}$) across the IRAC image 
to find the location of a peak is very similar to doing a sub-pixel centering.
Once $P_{decon}$ is found,
we subtracted a scaled PSF from the surrounding $1'\times1'$ area 
centered on $P_{decon}$.
The percentage of the subtracted flux (``CLEAN gain'') depends on $F_{AP9}$;
if $F_{AP9}$ is greater than $20\sigma$
\footnote{
The sigma here is the local background fluctuation 
measured using the following steps:
(a) running SExtractor in the single-image mode 
on all the IRAC mosaics and generating the segmentation maps;
(b) masking the detected sources according to the segmentation maps 
to generate the background images;
(c) convolving the background images with a 9$\times$9-pixel top-hat kernel 
and then generating the noise maps by calculating the rms around each pixel 
on the convolved background images.
}, then 1\% of the flux was subtracted;
if $F_{AP9}$ is less than $20\sigma$ but greater than $5\sigma$,
then 10\% of the flux was subtracted;
if $F_{AP9}$ is less than $5\sigma$, then 100\% of the flux was subtracted.
Our gain for brighter peaks (1\% for $20\sigma$) 
is much smaller than the values in most radio CLEAN (10\%).
We found a gain of 10\% for saturated or extended bright sources 
sometimes produces unreliable results.
This is likely because the IRAC PSF is not as well 
determined as the synthesized beam in radio imaging,
so that over- or under-subtraction easily occurs.
We found that a gain of 1\% produces a good balance 
between processing speed and deconvolution quality.
The subtracted flux ($F_{SUB}$) was summed and registered to 
the associated object in the $J$+$K_S$ segmentation map.
After each subtraction, 
the IRACLEAN process is repeated on the subtracted image,
until there were no pixels with $F_{AP9}$ higher than $2\sigma$.
This threshold is chosen such that 
it is not too low to allow for unreliable objects entering IRACLEAN, 
and it is not too high to cut off too much useful information on weak objects. 
We emphasize again that the concept of the deconvolution here 
is identical to that of CLEAN in radio imaging.

Under some extreme conditions
(unmatched PSF with saturated or bright extended objects), 
gains of $<1\%$ still do not work.
The residual fluxes of these objects would start to oscillate and diverge
with the progress of IRACLEAN,
and this phenomenon usually happens
when the residual fluxes are about 30\% to 50\% of the original fluxes.
Since the PSF area we used is $1'\times1'$,
this effect would also seriously affect 
the IRACLEAN results of other objects within the $1\arcmin$ area.
The IRACLEAN process will set a gain of 100\% 
to subtract the fluxes for such objects
when the signs of flux divergence show up,
and then stop to clean them hereafter.
It is worth noting that since greater than 50\% of the flux
is still cleaned by the normal IRACLEAN procedure,
the bias effect of assigning a gain of 100\% to the residual fluxes
should be much less than 50\% (i.e., $\ll$0.5 mag).

\subsubsection{Flux and Error Measurements}\label{IRACfluxerr}
At the end of the IRACLEAN process,
the summation of $F_{SUB}$ for each object is 
the flux measurement of that object,
and the final subtracted image is the residual map.
If an object cannot be detected in IRAC images
(i.e., its $F_{AP9}$ is lower than $2\sigma$), 
0.0 is assigned to its flux measurement.
The residual map allows us to check the quality of IRACLEAN,
and is also used for estimating flux errors.
The flux error of each object was calculated based on 
the fluctuations in the local area around that object in the residual map.
Any imperfection of the PSF would cause larger fluctuations
in the residual map,
and this effect is included in the flux error calculation.

\begin{figure}
\epsscale{0.9}
\plotone{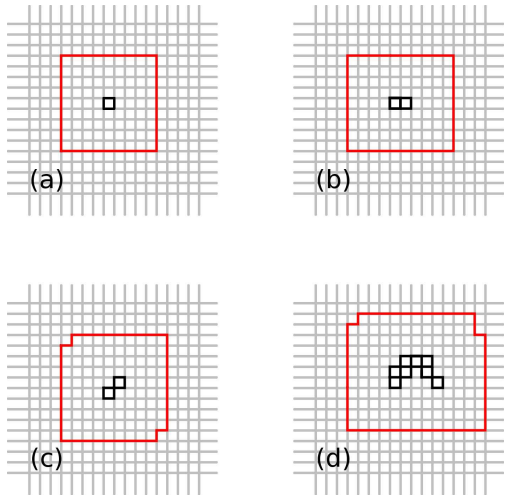}
\caption{Examples for calculating aperture sizes for various cases.
The grid in grayscale indicates the pixels in the resampled IRAC images;
the pixel marked with black color indicates $P_{decon}$;
and the area marked with red color shows the aperture size for this object.
Four different cases are shown in this plot:
(a) a bright point source
(only one $P_{decon}$ in the object center):
the aperture shape is a $9\times9$-pixel box,
i.e., the total number of pixels in the aperture is 81.
(b) a slightly extended source
(two nearby $P_{decon}$s around the object center):
the aperture shape is a $9\times10$-pixel rectangular,
and the total number of pixels in the aperture is 90.
(c) a slightly extended source with different position angle
as compared to case (b):
the aperture size is 98 pixels.
(d) an extended source:
the aperture shape is more irregular
and the aperture size is 140 pixels.
\label{apertureshape} }
\end{figure}

To estimate the flux error of an object,
we convolved the residual map with a $9\times9$-pixel top-hat kernel 
and then generated a noise map by calculating the rms 
around each pixel on the convolved residual map.
In conventional aperture photometry, the flux error scales 
with the square-root of the aperture area.
To follow this rule, 
we thus defined an effective aperture size for each object 
using the following method. 
First, if the IRACLEAN process always deconvolves an object
at the same $P_{decon}$ 
(i.e., deconvolving a bright point source with a matched PSF),
then the effective aperture size for this object is 81 pixels 
($9\times9$, Figure~\ref{apertureshape}a).
If the IRACLEAN process deconvolves a slightly extended object 
on two adjcent $P_{decon}$,
then the effective aperture size for this object can be 90 pixels 
(Figure~\ref{apertureshape}b), or 98 pixels (Figure~\ref{apertureshape}c).
For a more extended source, 
the distribution of these $P_{decon}$ can be more complicated, and
the shape of the aperture can be irregular (Figure~\ref{apertureshape}d).

Once the effective aperture size is determined, 
the flux error can be estimated with
\begin{eqnarray}
F_{err} = \sqrt{\frac{N_{pixel}}{81}} 
\times \frac{\sum\limits_{i=1}^{n}rms_{i}}{n},
\label{fluxerr}
\end{eqnarray}
where $N_{pixel}$ is the total number of pixels in the effective aperture,
$n$ is the total number of $P_{decon}$,
and $rms_i$ is the value from the noise map 
for a certain $P_{decon}$.
The first term in Equation~\ref{fluxerr} is 
to scale the flux error with the effective aperture size,
and the second term is the average flux error 
for the local area around this object.
One can use a more complicated method of calculating the second term 
by compute the local flux errors using different weightings.
For simplicity, we just use an unweighted mean,
which is similar to that used in conventional aperture photometry 
(cf., e.g., photometry with PSF fitting).
Using Equation~\ref{fluxerr}, we are able to reasonably estimate the flux error 
for all objects regardless of their morphologies.
For an object undetected in an IRAC image,
we used the coordinate of that object derived from the $J$+$K_S$ image,
and calculated its flux error as if it was a point source
(i.e., Using Equation~\ref{fluxerr} with $N_{pixel}$ = 81, $n$ = 1,
$rms$ = the value from the noise map for the $P_{decon}$,
where the $P_{decon}$ is the object coordinate in the $J$+$K_S$ image
provided by SExtractor).

Since we used a $9\times9$-pixel box 
as the flux aperture for IRACLEAN, 
an aperture correction needs to be applied for the flux errors.
The value of aperture correction is derived
from the deconvolution PSF for each channel.
We describe the aperture correction values in Section~\ref{7psf}.

\subsubsection{Position-Dependent PSF}\label{7psf}
The PSFs in the ECDFS IRAC images change with position. This is mainly caused
by the change in the orientation of \emph{Spitzer} during the observations.
In the GOODS-S area, the upper and lower half fields were observed 
with IRAC in two different epochs, with some overlap in the middle.
The orientation of $Spitzer$ changed by nearly 180 degrees between
the two epochs.
The SIMPLE data were also taken in two different epochs 
with a several-degree difference in $Spitzer$'s orientation. 
More than $80\%$ of the SIMPLE area is covered by both epochs.
In addition, there is also some overlap 
between the GOODS-S and SIMPLE observations.
Because of the asymmetric IRAC PSFs and different orientations of $Spitzer$
in the various observing epochs, 
the stacked IRAC image has several different synthesis PSFs.  
Using a single PSF to clean the entire image
will significantly degrade the IRACLEAN quality
because the performance of IRACLEAN critically depends
on the accuracy of the deconvolution PSF.
IRACLEAN is affected by the accuracy of the deconvolution PSF in two aspects: 
(a) For a bright point source, 
using an inaccurate PSF would lead to distorted intermediate residual images 
after iterating the clean process many times. 
From this point, 
the local brightest pixel (i.e., $P_{decon}$) may not be 
around the center of this object 
so that IRACLEAN starts treating it as an extended source 
until the end of the clean process. 
This effect would lead to a biased flux measurement. 
Same thing would happen to bright extended sources, too. 
Fainter sources are not affected by this effect 
because their corresponding segmentation areas are only a few pixels 
hence their $P_{decon}$ pixels cannot go too far from the object centers. 
(b) Some fainter sources are, however, affected by another issue 
because of using an inaccurate PSF. 
If they are close to another objects that are not cleaned using accurate PSFs, 
then the dirty residuals of the wings of their close neighbors 
would contanminate their flux measurements. 
The brighter the close neighbor is, the more serious the effect is; 
the closer the separation is, the more serious the effect is.
In order to obtain better flux measurements for all objects,
we used several different PSFs to clean one IRAC image.

\begin{figure*}
\epsscale{0.5}
\plotone{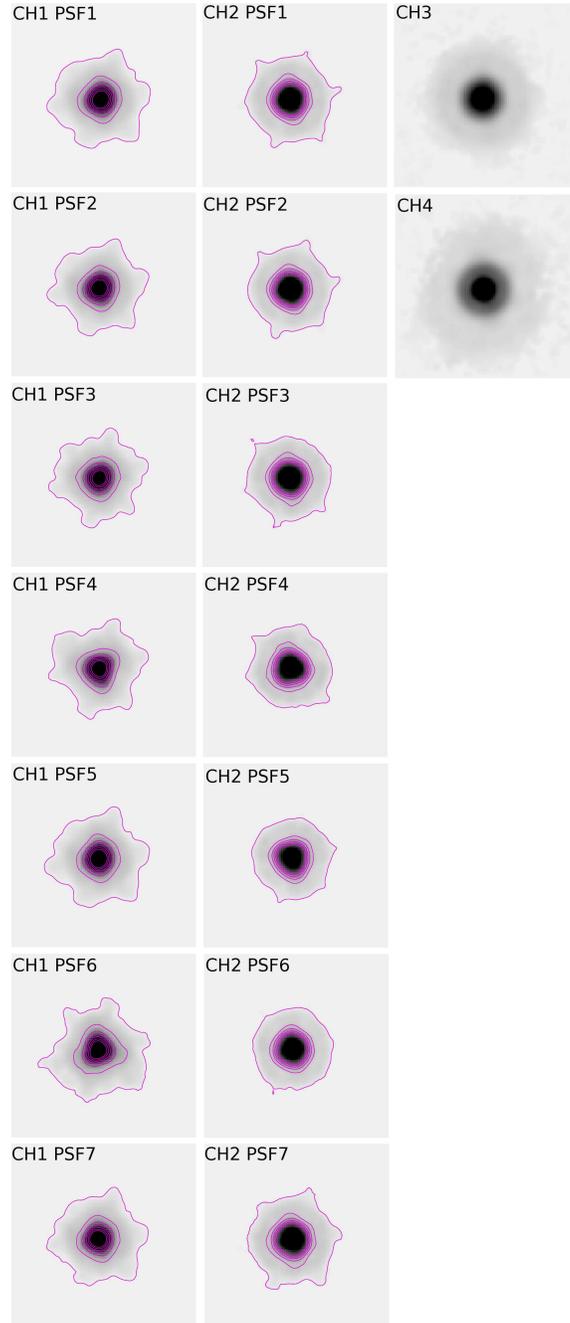}
\caption{The PSF images for IRAC 3.6$\mu$m to 8.0$\mu$m.
The image size is $1'\times1'$ in each panel.
There are seven testing PSFs for 3.6$\mu$m and for 4.5$\mu$m,
and one PSF for 5.8$\mu$m and 8.0$\mu$m.
Contours are plotted on the PSF images for IRAC 3.6$\mu$m and 4.5$\mu$m
to make the differences between PSFs visually clearer.
\label{PSFimage} }
\end{figure*}

It is possible to construct many different PSFs in an IRAC image.
We start with as many as seven different PSFs but below we will demonstrate 
that only four PSFs are necessary for 3.6 and 4.5 $\mu$m. 
First, we generated seven PSFs by picking up bright point-like sources
from seven different areas in the IRAC image for each IRAC band.
The generated PSFs are shown in Figure~\ref{PSFimage}.
Then for a given IRAC band, we repeated the IRACLEAN process seven times 
using one of the seven PSFs each time. 
This gives seven flux measurements for each object.
Since our flux errors (Section~\ref{IRACfluxerr}) contain 
the effect of unmatched PSF,
we can use the flux errors as an indicator on
which PSF works the best on each object.

\begin{figure*}
\epsscale{1.0}
\plotone{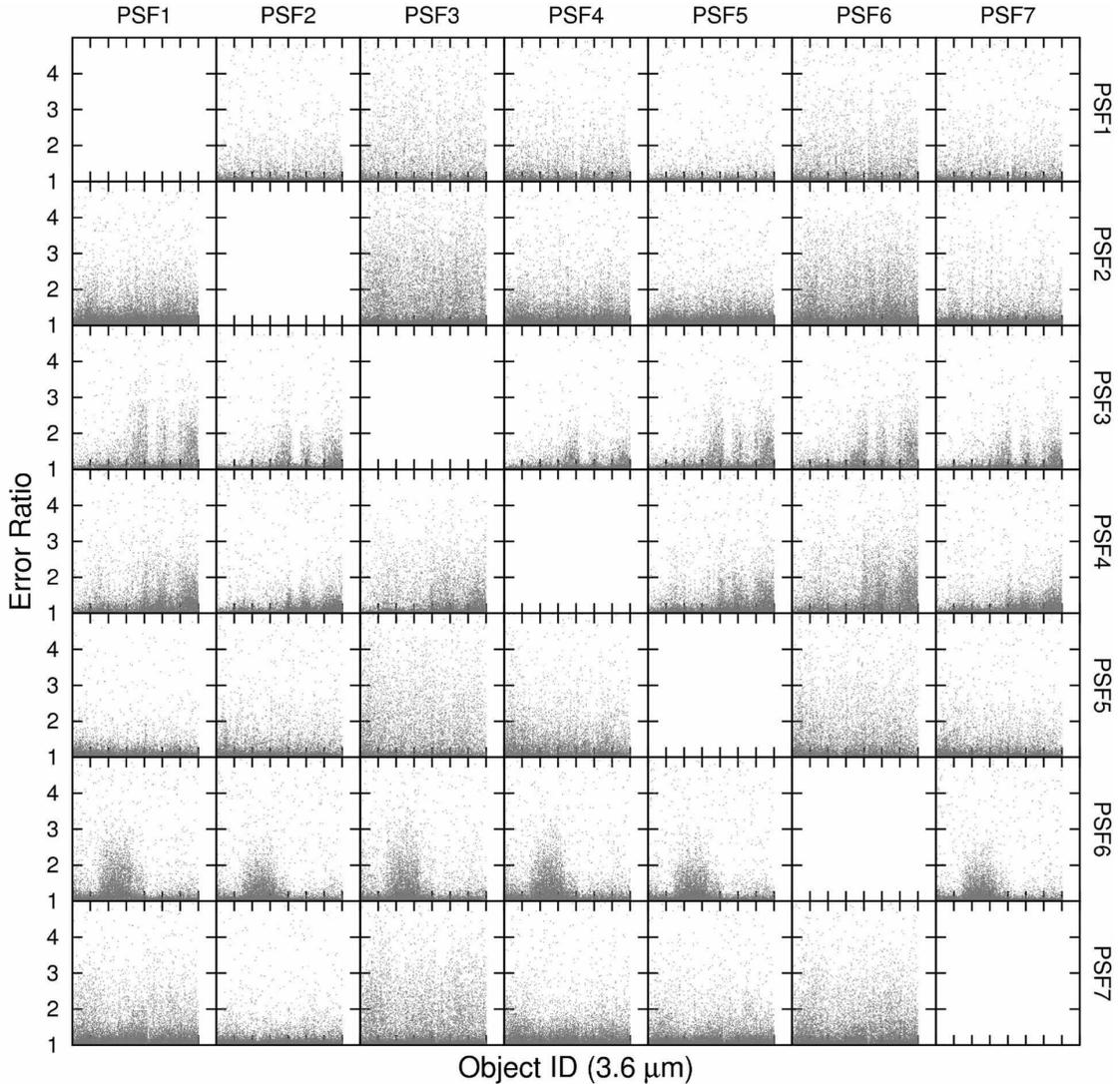}
\caption{Comparison of the performance of the seven PSFs 
at 3.6 $\mu$m.  The $i$-th column contains objects whose flux errors are the 
lowest when IRACLEANed with the $i$-th PSF.  The $j$-th row compares the
errors derived with the $j$-th PSF against the errors 
with the $i$-th (best) PSF.
X-axis is object ID and Y-axis is error ratio in each panel.
See text for details.
\label{errdiff1} }
\end{figure*}

\begin{figure*}
\epsscale{1.0}
\plotone{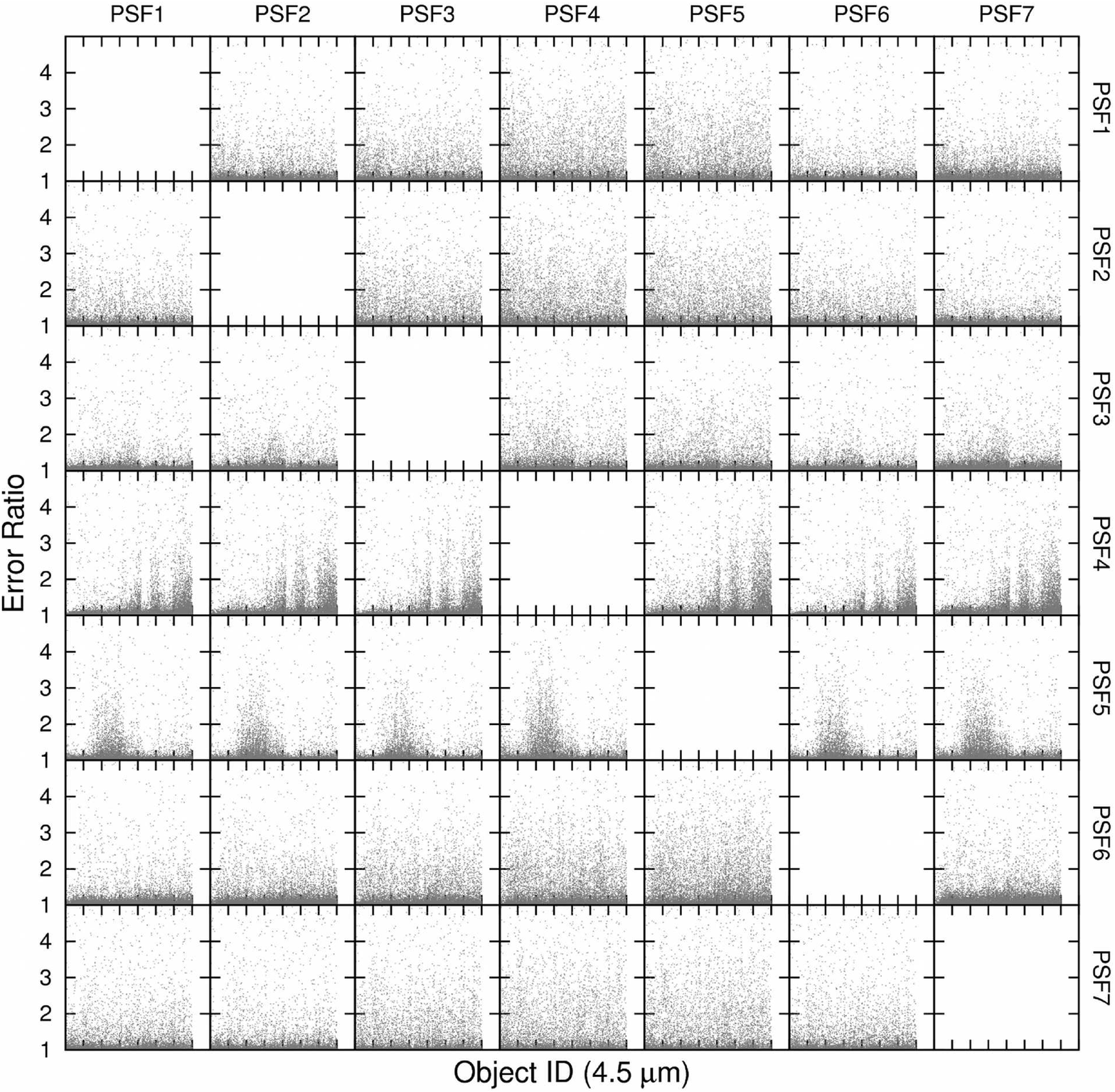}
\caption{Same as Figure~\ref{errdiff1} but for 4.5$\mu$m.
\label{errdiff2} }
\end{figure*}

Figures~\ref{errdiff1} and \ref{errdiff2} compare the flux errors 
based on the seven PSFs at 3.6 and 4.5 $\mu$m, respectively. 
In the $i$-th column, 
we pick up objects whose flux errors are the lowest when IRACLEANed with 
the $i$-th PSF. Then in the $j$-th row, we compare their flux errors based on
the $j$-th PSF with respect to the flux errors based on the $i$-th (best) PSF. 
The x-axis in each panel is object ID 
and the y-axis is the error ratio of $j$-th to $i$-th PSF.
(Since in each panel we are always comparing against the best PSF,
the flux ratios in all the panels are always greater than 1.) 
In other words, the comparison shows how the other PSFs perform 
when compared to the best-matched PSF for each object.

By examining Figures~\ref{errdiff1} and \ref{errdiff2},
we can determine which PSF is necessary and which PSF is redundant.
In a panel, if most of the error ratios are fairly close to 1,
it means that the IRACLEAN results using the $i$-th PSF and the $j$-th PSF
have similar quality.
In such a case, the $i$-th and $j$-th PSFs can be replaced by each other.
For example, we found that for 3.6 $\mu$m, PSF5 and PSF7 can both be replaced
by PSF2, and PSF1 can be replaced by any of the other PSFs.
In other words, only PSF2, 3, 4, and 6 are necessary for IRACLEAN at 3.6 $\mu$m.
Similarly, only PSF1, 2, 4, and 5 are necessary for 4.5 $\mu$m.
For 5.8 and 8.0 $\mu$m, 
we do not find significant differences among the seven PSFs,
so only one PSF each is adopted and is shown in Figure~\ref{PSFimage}.

\begin{figure*}
\epsscale{1.0}
\plotone{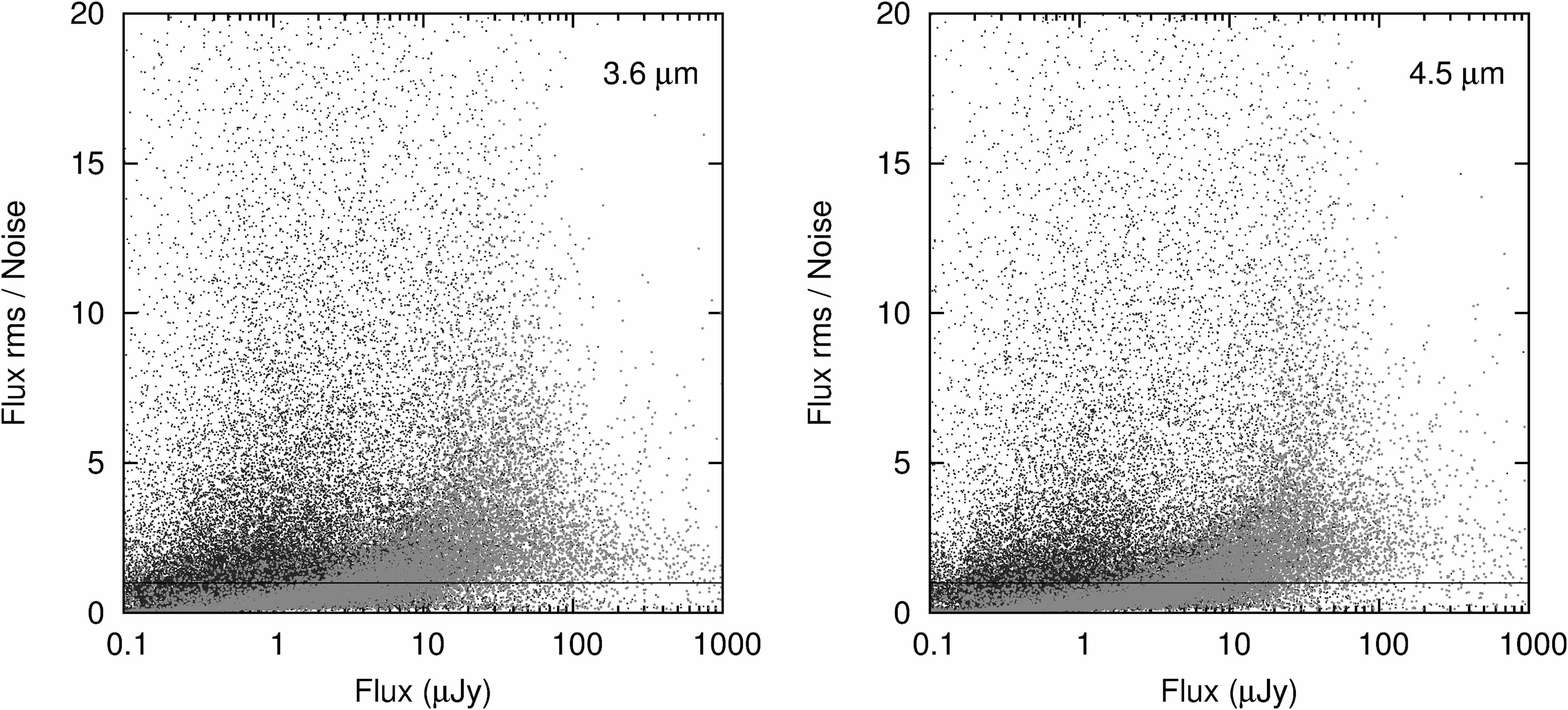}
\caption{IRACLEAN performance affected by the accuracy of the deconvolution PSF.
The left panel is for IRAC 3.6 $\mu$m
and the right panel is for IRAC 4.5 $\mu$m.
The Y-axis is the ratio between the scatter of fluxes measured
using different PSFs and the residual noise.
The X-axis is the IRAC flux measured using the best PSF.
The black solid line indicates a ratio of 1.0 for Y-axis
The higher the flux rms to noise ratio is, the more serious the PSF effect is.
The light grey data points indicate isolated objects defined as that
they have no neighbors brighter than 30\% of their fluxes within $6\farcs0$.
The dark grey data points are the rest of objects.
The ratio between the numbers of isolated and non-isolated objects is about 2:5.
The IRACLEAN fluxes of the isolated and non-isolated objects
can be affected by the accuracy of the deconvolution PSF
because of different reasons. See text for details.
\label{fluxrms} }
\end{figure*}

To show how seriously IRACLEAN is affected by 
the accuracy of the deconvolution PSF, 
we calculated the rms scatters of the IRACLEAN measurements 
using the seven different PSFs for all the objects in 3.6 $\mu$m and 4.5 $\mu$m,
and compared them with the noises measured from the residual images. 
The results are shown in Figure~\ref{fluxrms}.  
According to Figure~\ref{fluxrms}, 
the IRACLEAN fluxes of the bright isolated objects are affected 
by PSF more seriously than that of the faint isolated objects, 
which can be explained by the abovementioned effect (a). 
Most of non-isolated objects are faint objects 
since bright objects have lower surface number density 
hence they are more isolated. 
These non-isolated faint objects show another distribution in this plot. 
Many of them are also seriously affected by inaccurate PSF,
which can be explained by the abovementioned effect (b).
Figure 14 proves that the performance of IRACLEAN critically depends on the PSF,
and using multiple PSFs to clean the IRAC 3.6 $\mu$m and 4.5 $\mu$m images 
can gain the most advantage from IRACLEAN.

With the above experiments,
we adopted the IRACLEAN results 
based on the final four PSFs for 3.6 and 4.5 $\mu$m,
and one PSF each for 5.8 and 8.0 $\mu$m.
For a given object, we pick the flux measurement with the lowest error among
the four PSFs at 3.6 and 4.5 $\mu$m.
We then calculated the aperture correction factors 
for each channel based on the PSFs.
For 3.6 $\mu$m and 4.5 $\mu$m, 
the correction factors derived from the four PSFs agree with each other
within 1\% and we adopted the mean values.
It is worth noting that 
the correction factors from different PSFs are very similar
because these PSFs are synthesized from the same PSF 
with different orientations,
hence the ratios of fluxes between inside and outside the aperture 
for different PSFs agree with each other very well.
It also suggests that the centering issue of generating the stacking PSFs
is negligible.
The adopted aperture correction factors are
1.894, 1.875, 2.111, and 2.334 for 3.6, 4.5, 5.8, and 8.0 $\mu$m, 
respectively, for our $2\farcs7 \times 2\farcs7$ flux aperture.

\subsubsection{Known Issues}\label{issues}
Three extremely bright (thus saturated) stars and several very bright
objects are in the ECDFS.
They have very bright wings and occupy areas 
that are much broader than $1'\times1'$ (the deconvolution PSF size).
Objects close to these bright sources, 
including those around the diffraction spikes and crosstalk features,
therefore have higher flux measurements.
In addition, as mentioned in Section~\ref{clean},
we stop the IRACLEAN process if it starts diverging for some bright objects.
This also implies that the flux measurements of nearby objects are affected.
However, we should note that the flux measurements of these nearby objects
will be affected even more seriously 
if we do not stop the IRACLEAN process when divergence happens.

\subsection{Quality and Performance}

\subsubsection{Residual Images}
\begin{figure*}
\epsscale{0.9}
\plotone{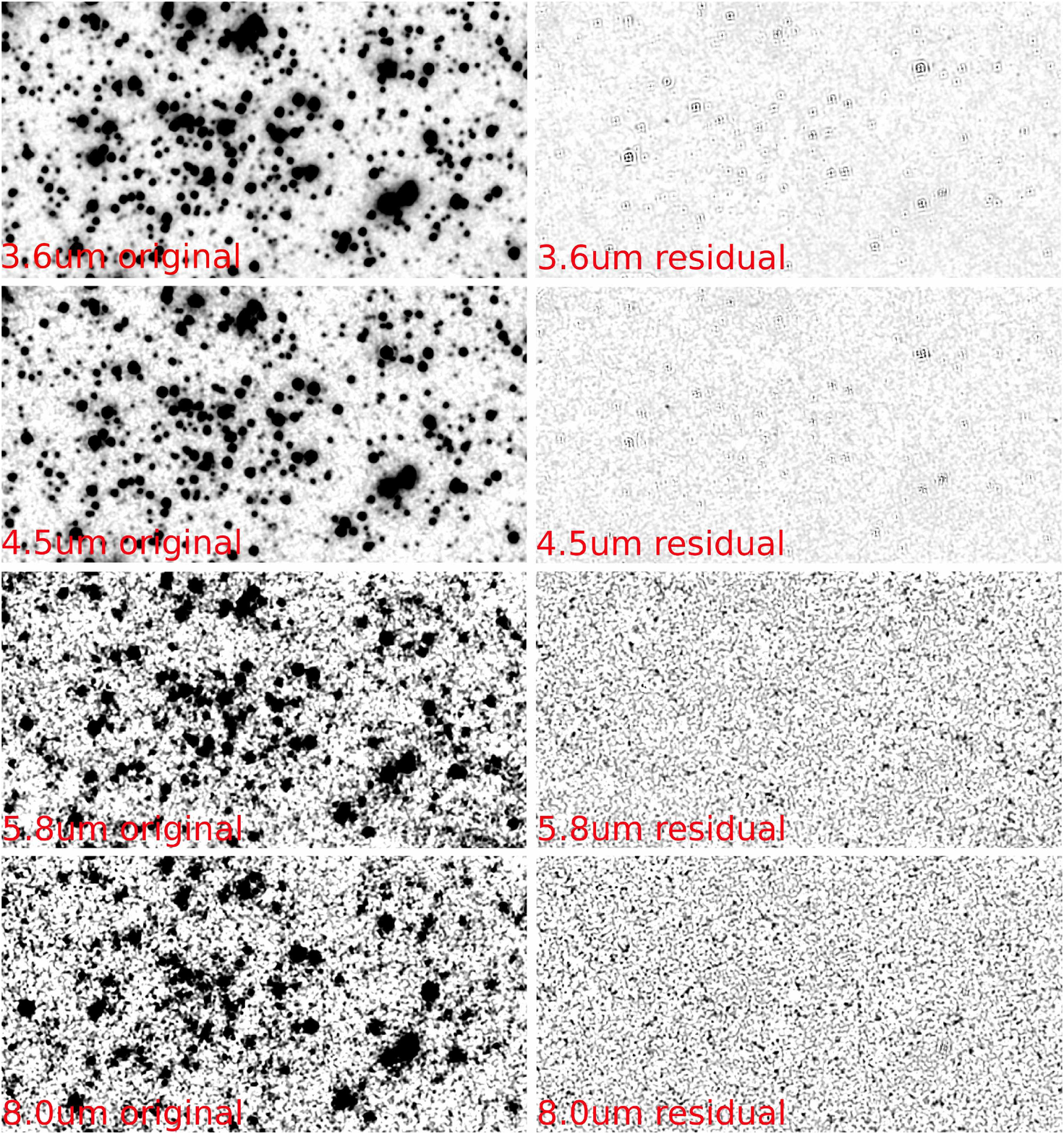}
\caption{Performance of IRACLEAN.
The SIMPLE IRAC images are shown in the left panels and
their residual images are in the right panels.
The region in each panel is $300''\times140''$.
North is up and east is to the left.
The brightness and contrast of each panel are identical.
From top are the IRAC images of 3.6$\mu$m (channel 1) to 8.0$\mu$m (channel 4).
All images are shown with inverted linear scales.
\label{irac-decon} }
\end{figure*}

We demonstrate the performance of IRACLEAN in Figure~\ref{irac-decon}.
Regions of sizes $300''\times140''$ in the SIMPLE IRAC images (left panels)
and their residual images (right panels) are shown.
All images are shown with inverted linear scales.
The brightness and contrast of each panel are identical.
By visually checking the residual images,
we found that IRACLEAN works reasonably well in all four channels.
For 3.6$\mu$m and 4.5$\mu$m, however, there are halos around many sources 
in the residual images,
which may be due to unmatched deconvolution PSFs and/or
under-sampled PSF caused by the large pixel scale of IRAC.
The 3.6 $\mu$m or 4.5 $\mu$m residual images shown in Figure~\ref{irac-decon}
are generated using a single deconvolution PSF.
However, as discussed in Section~\ref{7psf}, there are indeed
four PSFs for each band, 
and thus three more versions of residual images each, 
which are not shown here.
Since we adopt the result 
that has the lowest residual fluctuation for each object,
the IRAC fluxes for 3.6$\mu$m and 4.5$\mu$m in our photometric catalog
are indeed better than the visual impression based on any single residual image.
On the other hand, for 5.8$\mu$m and 8.0$\mu$m,
the residual maps are very clean;
there are no obvious residual effects or artifacts.

\subsubsection{Monte-Carlo simulations}\label{montecarlo}
To further understand the performance of IRACLEAN,
we carried out simple Monte-Carlo simulations
for objects with high S/N (S/N$\sim50$) blended by nearby objects.
All the simulated objects have flat spectra and are point-like.
The separations between the blended objects 
are from $1\farcs2$ to $3\farcs9$ 
and the flux ratios between the blended objects are from 1 to 100.
We simulated a $J$+$K_S$ image of the blended objects and corresponding images 
for four IRAC channels.
The PSF of the $J$+$K_S$ image is extracted from the real $J$+$K_S$ data.
The PSF for each object in the simulated $3.6\mu$m and $4.5\mu$m images
is randomly picked up from the seven PSFs
described in Section~\ref{7psf} and in Figure~\ref{PSFimage}.
We ran SExtractor on the simulated $J$+$K_S$ images
to generate the segmentation map,
and then performed IRACLEAN on the simulated IRAC images.
We adopted only four PSFs for IRACLEAN for 3.6$\mu$m and for 4.5$\mu$m,
as in our real IRACLEAN (see Section~\ref{7psf}).

\begin{figure*}
\epsscale{0.7}
\plotone{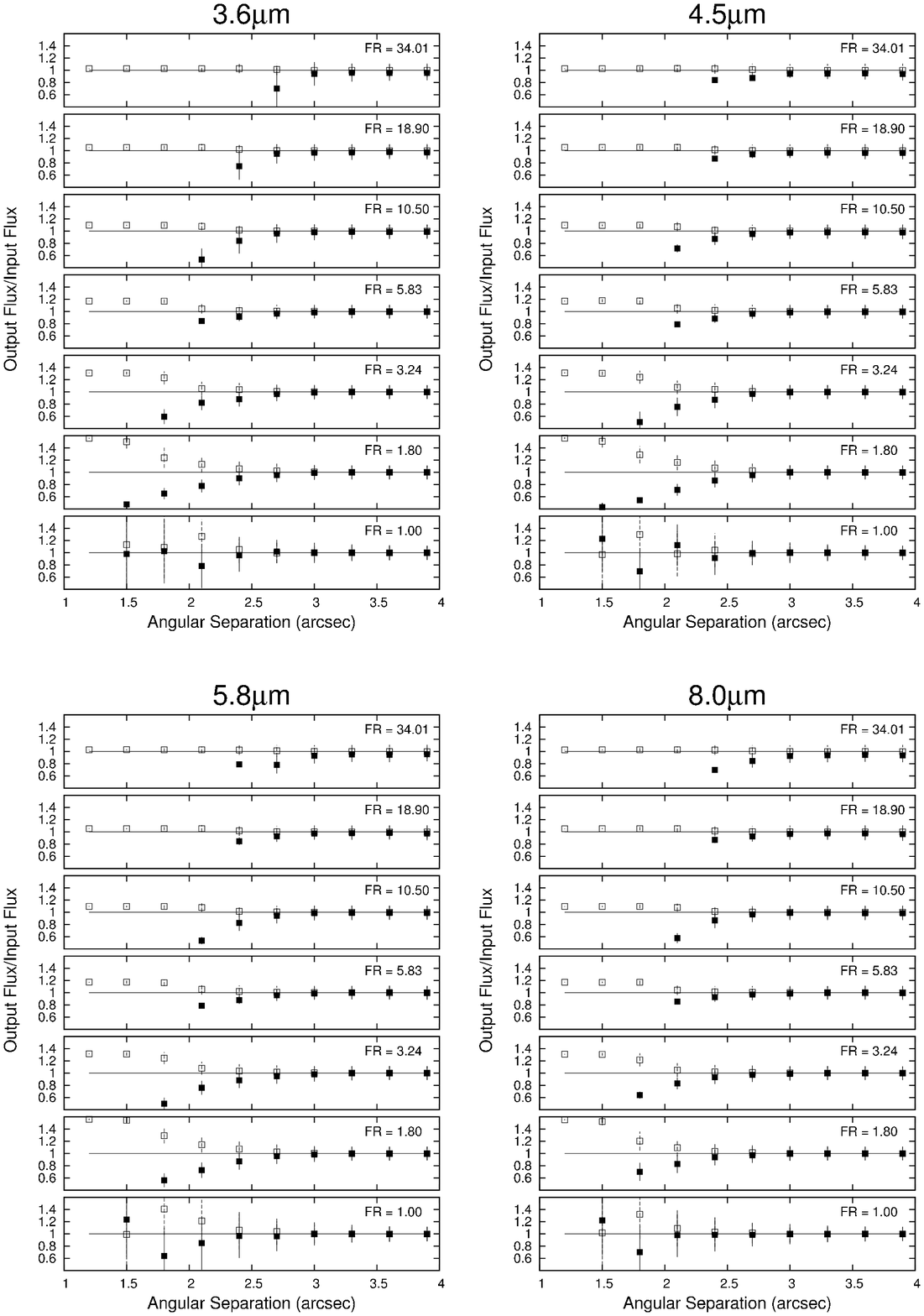}
\caption{Simple Monte-Carlo simulations of IRACLEAN for two blended objects.
The simulation includes different angular separations 
and flux ratios between two blended objects.
Four large panels indicate the simulation results for four IRAC channels.
The FR value shown in each sub-panel is the flux ratio 
between two blended objects.
The ratios between the output flux and the input flux of objects
in the blended pairs are shown in the vertical axes.
The open box indicates the brighter object in a pair 
while the filled box indicates the fainter one.
Some missing filled boxes in each panel mean that
SExtractor cannot successfully deblend the two objects in the WIRCam image;
IRACLEAN is not able to handle this case since it relies on SExtractor
for object detections.
The IRACLEAN fluxes of the brighter objects in pairs therefore
are the combined fluxes of both objects in pairs.
\label{monte} }
\end{figure*}

Figure~\ref{monte} shows the results of the Monte-Carlo simulations.
For all pairs with flux ratios $>40$ and some pairs 
with separations $<2\farcs5$, 
SExtractor cannot resolve them in the simulated $J$+$K_S$ image. 
On such cases, IRACLEAN does not know there exist two objects, 
and thus we have no handle on them. 
These cases are not shown in Figure~\ref{monte}. 
This is a limit of the WIRCam imaging, not IRACLEAN.
On the other hand, when SExtractor is able to resolve the objects, 
IRACLEAN has very good performance for blended objects 
with separations greater than $3\farcs0$ within the entire flux ratio range.
For the fainter objects in pairs,
the fluxes can be under-estimated by up to 50\% 
with separations less than $3\farcs0$;
the smaller the separation, the more severe the under-estimate is,
and it is flux ratio dependent.
For the bright objects in pairs, however,
the fluxes can be over-estimated by up to 50\%
with separations less than $2\farcs0$ when flux ratios are less than 3.5.
In these extreme cases, 
we are pushing both IRACLEAN and SExtractor to their limits.

To estimate how many sources in our catalog may suffer from 
the under-estimate issue of the fainter source in a pair,
we used the results of the Monte-Carlo simulations to determine
the flux ratio ($FR$) vs. angular separation relation
where the flux is under-estimated by 20\%,
and the relation can be described by
\begin{eqnarray}
AS = 1.84 + 0.5 \times log(FR),
\label{ASFR}
\end{eqnarray}
where $AS$ is the angular separation and $FR$ is the flux ratio.
In other words, if one pair with a flux ratio of $FR$
has an angular separation less than $AS$,
the flux of the fainter object in the pair is under-estimated 
by greater than 20\% (but not by $\gg$50\% according to our simulations).
We then used Equation~\ref{ASFR} to estimate 
how many sources in our catalog are in the abovementioned situation.
The numbers are 3347 (5.4\%), 2840 (4.6\%), 1086 (1.7\%), and 691 (1.1\%)
for 3.6$\mu$m, 4.5$\mu$m, 5.8$\mu$m, and 8.0$\mu$m, respectively.
Fewer sources are effected in 5.8$\mu$m and 8.0$\mu$m
because of their much lower surface number densities 
due to the shallower detection limits (see Section~\ref{depth} for details).

To estimate how many sources in our catalog may suffer from 
the over-estimate issue of the brighter object in a pair,
we utilized Equation~\ref{ASFR} with $FR\leq3.5$ 
to repeat the counting but for the brighter sources.
The numbers are 872 (1.4\%), 707 (1.1\%), 430 (0.7\%), and 272 (0.4\%)
for 3.6$\mu$m, 4.5$\mu$m, 5.8$\mu$m, and 8.0$\mu$m, respectively.
It is worth noting that the real number of objects 
which are affected by the under-estimate and over-estimate issues
may be less than the abovementioned estimated number
because the $FR$s calculated using the fluxes of real objects in our catalog
are already biased by the under-estimate and over-estimate issues
so that the resulting $AS$' from Equation~\ref{ASFR} are greater than
the ideal one calculated using the unbiased fluxes.

These results suggest that the performance of IRACLEAN is generally good
and only a small number of objects in our catalog 
are affected by their neighbors, and the effect is not $\gg$50\%.
This simple simulations, however, do not include
different morphologies, number of blended objects, and S/N ratios.
We therefore perform another simulation using the mock IRAC images
to take the abovementioned parameters into account,
and show the results in Section~\ref{iracmock}.

\subsubsection{Mock IRAC image simulations}\label{iracmock}
\begin{figure*}
\epsscale{0.9}
\plotone{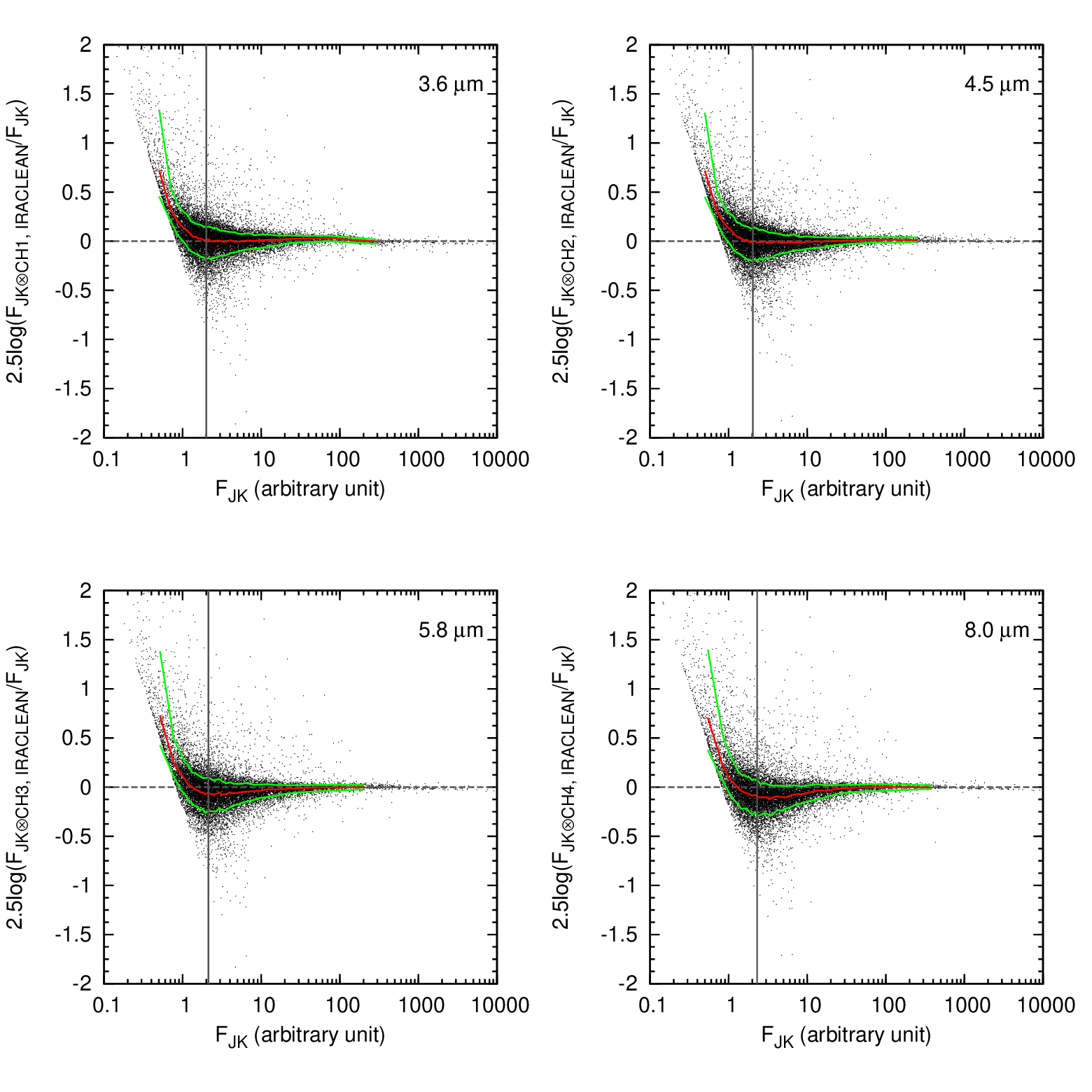}
\caption{IRACLEAN performance on the mock IRAC images.
Four panels from left to right, top to bottom
are for 3.6 $\mu$m, 4.5 $\mu$m, 5.8 $\mu$m, and 8.0 $\mu$m.
The Y-axis is the magnitude difference between
the IRACLEAN flux in the mock IRAC image 
and the SExtractor flux in the original $J$+$K_S$ image.
The X-axis is the SExtractor flux in the original $J$+$K_S$ image.
The red line indicates the running median
while the green lines indicate 
the upper and lower 68th percentiles of the distribution.
The vertical grey line indicates the 5$\sigma$ limit of IRACLEAN flux.
\label{iracleanmock} }
\end{figure*}

To evaluate the IRACLEAN performance affected by
the effects of morphology, number of blended objects, and S/N ratio,
We generated four mock IRAC images by convolving the $J$+$K_S$ mosaic
with the ideal IRAC PSFs for the four IRAC channels.
We then ran IRACLEAN on these mock IRAC images
with the segmentation map of the original $J$+$K_S$ image.
Assuming the SExtractor FLUX\_AUTO measurement of each object 
in the original $J$+$K_S$ image is the true flux amplitude
\footnote{
According to the SExtractor flags in the $J$+$K_S$ catalog, 
about 70\% of the total number of objects have significantly biased FLUX\_AUTO
because of bright and close neighbors, 
and/or were originally blended with another one.
Therefore the FLUX\_AUTO measurements for these cases
many not be the true answer.
However, the mock IRAC images have even more severe blending issue
as compared to the original $J$+$K_S$ image.
The assumption here therefore is still reasonable and feasible.},
we can examine how well IRACLEAN performs under a complicated condition
by checking if IRACLEAN can recover the SExtractor fluxes.
The results are shown in Figure~\ref{iracleanmock}.
For 3.6 $\mu$m and 4.5 $\mu$m, 
the fluxes in general are not under-estimated 
according to their running medians.
For 5.8 $\mu$m and 8.0 $\mu$m, however,
the fluxes around 3.0 are under-estimated by 0.05 to 0.1 mag in general
because of their larger PSFs.
We should note that
for the real SIMPLE IRAC data, the depths in 5.8 $\mu$m and 8.0 $\mu$m
are much shallower than that in $J$, $K_S$, 3.6 $\mu$m, and 4.5 $\mu$m,
so the surface number densities in 5.8 $\mu$m and 8.0 $\mu$m
are much lower than that in the other bands.
The under-estimated issue for 5.8 $\mu$m and 8.0 $\mu$m shown 
in our simulations should therefore be much milder in the real SIMPLE data.
For fluxes fainter than 2.0, they are over-estimated
because of the selection effect in the fainter data.
According to this plot,
we conclude that the real SIMPLE IRACLEAN fluxes 
with greater than $5\sigma$ detections would have biases $\ll$0.1 mag.

\begin{figure*}
\epsscale{0.9}
\plotone{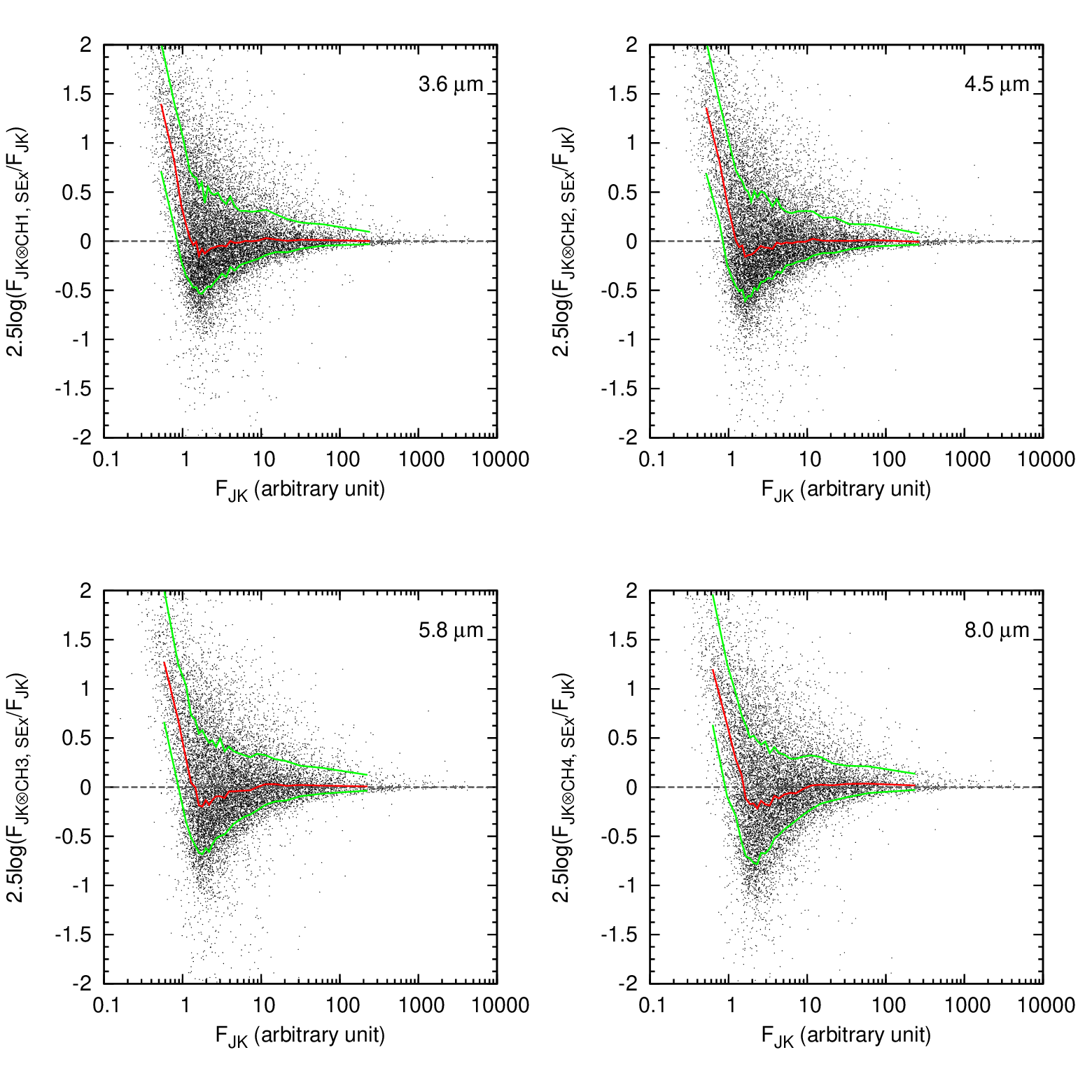}
\caption{Same as Figure~\ref{iracleanmock} but using SExtractor.
\label{sexmock} }
\end{figure*}

We also ran SExtractor in the single-image mode for the mock IRAC images
and the results are shown in Figure~\ref{sexmock}.
According to Figure~\ref{sexmock},
the under-estimate issue of SExtractor is more severe than that of IRACLEAN,
and the scatters are at least a factor of 2 of that using IRACLEAN.
The results suggest that IRACLEAN should perform better than SExtractor
for the SIMPLE IRAC images.

\subsubsection{Comparison with the cross-convolution method}
In \citet{wang2010}, an alternative cross-convolution method (XCONV)
was introduced for measuring $K_S$--IRAC color.
The concept of this cross-convolution method is
to match the PSF between $K_S$ and each IRAC channel
by convolving their PSFs to each other.
Since this method does not require any IRACLEAN procedure
but just simple photometric measurements on PSF-matched images,
it should provide the most accurate $K_S$--IRAC colors
for isolated bright objects. We thus compare our IRACLEAN results
with XCONV colors.

We first convolved each IRAC image with the WIRCam $K_S$ PSF, 
and convolved the WIRCam $K_S$ image with the IRAC PSFs we used in IRACLEAN.
We take into account that there exist four different IRAC PSFs 
at 3.6 and 4.5 $\mu$m (Section~\ref{7psf}).
The FWHMs of the XCONVed PSFs are between $2''$ to $3''$.
We then used the double-image mode of SExtractor
to measure fluxes on these XCONVed images,
by using the unconvolved $J$+$K_S$ image as the detection image.
Because we focus on measuring $K_S$--IRAC colors here,
we do not need to recover total fluxes
since we have matched the PSFs in the $K_S$ and IRAC images. 
We adopt an aperture size of $3''$ in diameter in all cases,
to have a good balance between S/N and 
the inclusion of most fluxes for all kinds of source morphologies.
These are all similar to the XCONV method in \citet{wang2010}.

\begin{figure*}
\epsscale{1.0}
\plotone{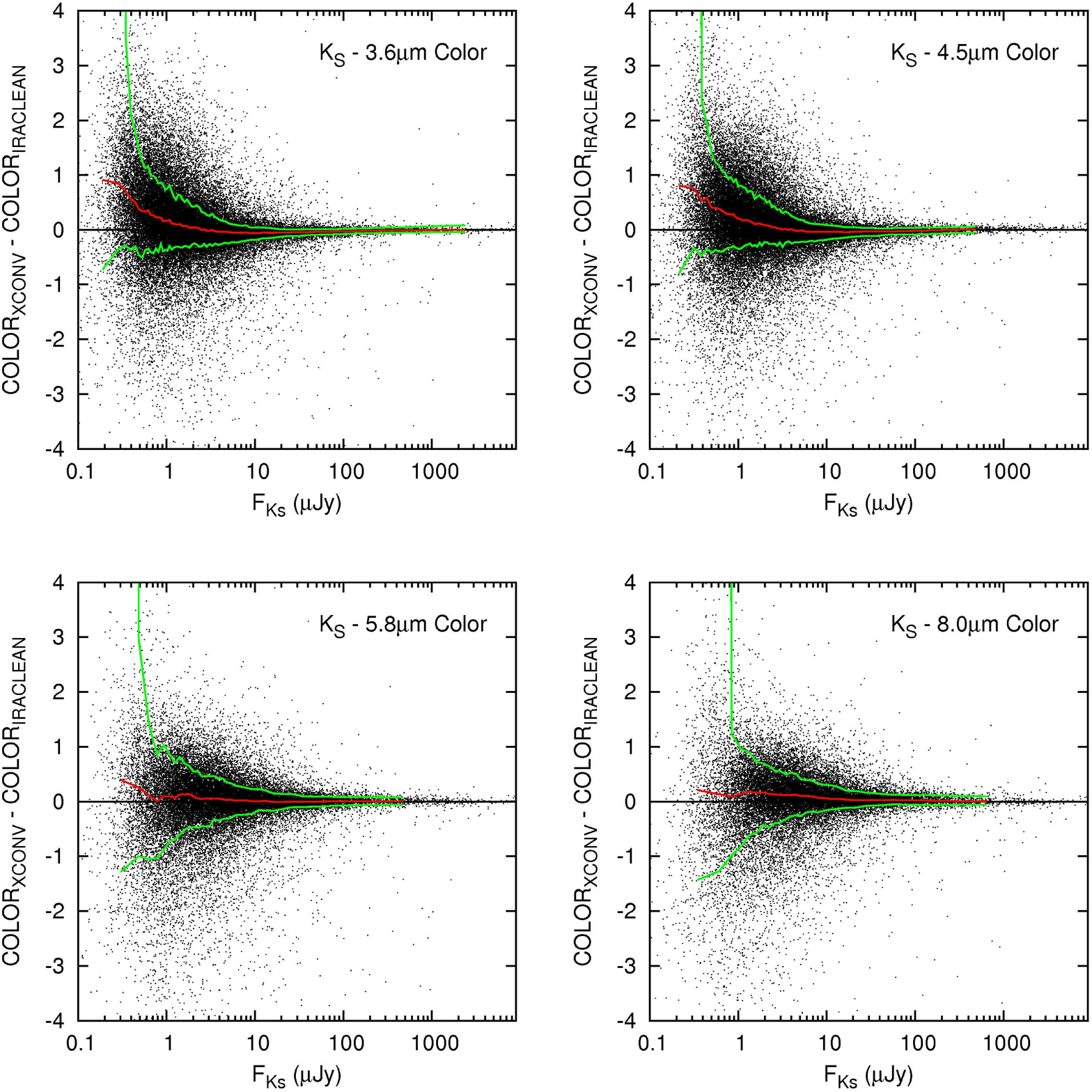}
\caption{Color differences between the IRACLEAN and XCONV methods.
The red line indicates the running median
and the green lines indicate the upper and lower 68th percentiles.
The systematic offsets between these two colors for 1,231 objects
with $F_{Ks} > 100$ are -0.032, -0.011, -0.031, and -0.011 
for 3.6$\mu$m, 4.5$\mu$m, 5.8$\mu$m, and 8.0$\mu$m, respectively.
The plot shows that the $K_S$ - IRAC colors derived using both methods
are consistent with each other.
\label{color_xconv} }
\end{figure*}

The comparisons of the $K_S$--IRAC colors derived using
the IRACLEAN and XCONV methods are shown in Figure~\ref{color_xconv}.
The $K_S$--IRAC colors derived using both methods
are consistent with each other.
The systematic offsets between these two colors (XCONV--IRACLEAN)
for 1,231 objects with $F_{Ks} > 100 \mu$Jy 
are -0.032, -0.011, -0.031, and -0.011 mag,
for 3.6, 4.5, 5.8, and 8.0 $\mu$m, respectively,
which are within their statistical errors.
We note that in the IRACLEAN case,
the $K_S$ fluxes are measured using the SExtractor AUTO aperture,
which can miss up to 5\% of total fluxes \citep{ba1996}.
The IRACLEAN measurements of the IRAC fluxes, 
however, does not have such systematics.
This may explain why the IRACLEAN colors are systematically 
redder than the XCONV colors. 
Nevertheless, the consistency between the two colors is fairly good.
We also checked the objects with color differences greater than 0.2 mag
and $F_{Ks} > 100 \mu$Jy.
Most of them are in crowded areas with multiple bright sources
so that their XCONV colors may be biased because of the large XCONV PSFs.
Only a few of them are saturated/bright extended objects
where their flux measurements are affected 
by the limitations of IRACLEAN, as we mentioned in Section~\ref{issues}.

\subsubsection{Comparison with SExtractor}\label{comp_sex}
\begin{figure*}
\epsscale{1.0}
\plotone{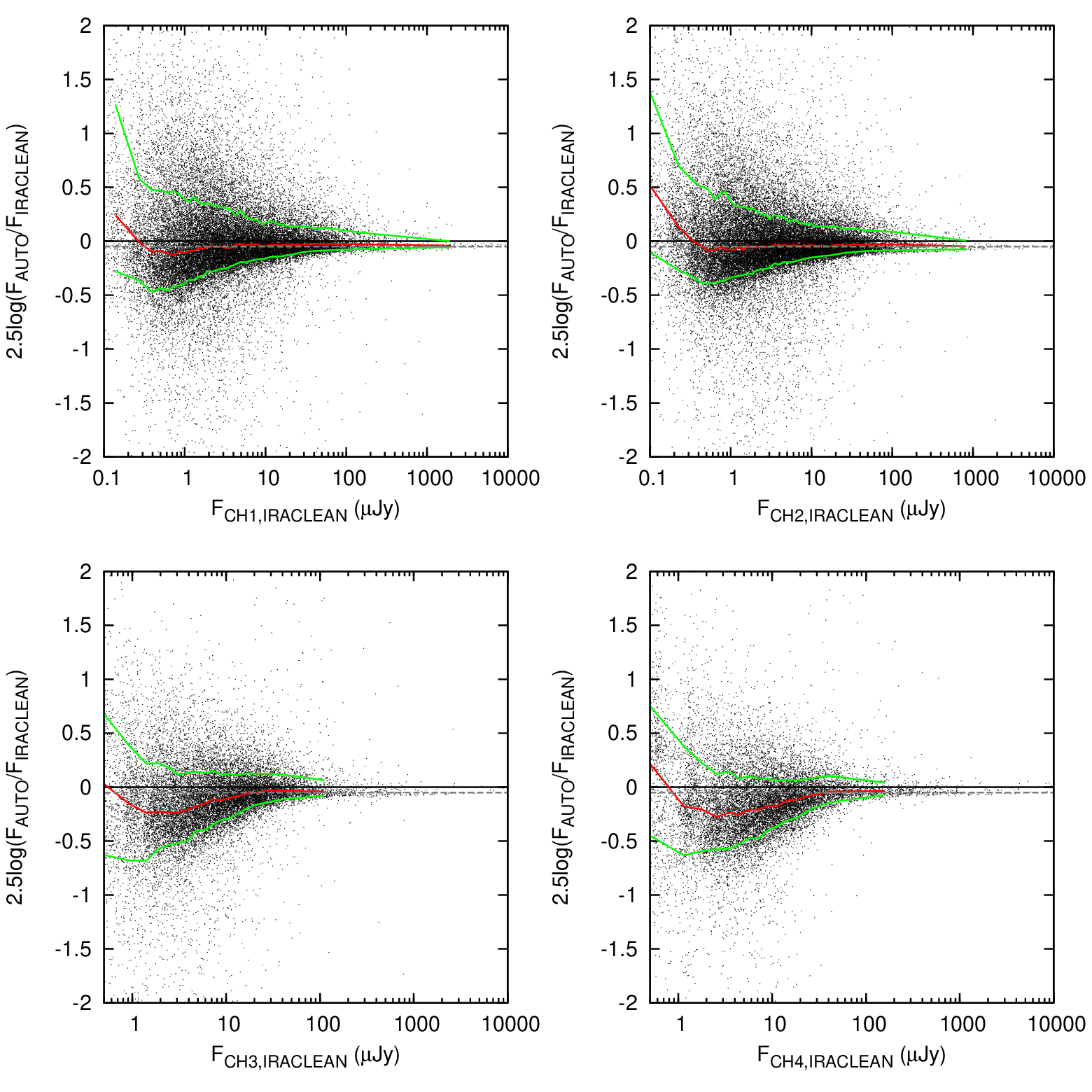}
\caption{The comparison of the IRAC photometry between the IRACLEAN and
FLUX\_AUTO methods.
X-axis is the flux for each IRAC channel
and Y-axis is the magnitude difference (or flux ratio)
between IRACLEAN and FLUX\_AUTO methods.
The black line indicates 0 offset
while the gray dashed line indicates an offset of -0.05 for Y-axis.
The red line indicates the running median
and the green lines indicate the upper and lower 68th percentiles.
The offsets between the results derived using the two methods
are consistent with the missing flux ratio provided by \citet{ba1996}.
\label{irac-comp} }
\end{figure*}

Figure~\ref{irac-comp} shows a comparison of 
IRAC photometry using IRACLEAN against
that using SExtractor with FLUX\_AUTO in the single-image mode.
The two photometric methods are consistent on bright objects, 
but the scatter increases on fainter objects.
Even on bright objects, as well as faint objects, 
there are small magnitude offsets.
The gray dashed lines in the figure indicate a magnitude difference of
-0.05, which well describe the differences 
in magnitudes derived using the two methods.
This is consistent with the 5\% flux loss of SExtractor FLUX\_AUTO 
discussed in \citet{ba1996}.

At 3.6 and 4.5 $\mu$m, the distributions of flux ratios 
are asymmetric about -0.05 mag.  
There are more objects whose SExtractor fluxes are larger than IRACLEAN fluxes. 
This is consistent with SExtractor fluxes being boosted by nearby objects.
Such a trend is not apparent at 5.8 $\mu$m and is even reversed at 8.0 $\mu$m.
In these two bands, the sensitivity is lower and thus
the mean separation between detected objects is larger.
This makes flux booting by nearby objects less an issue.
In addition, the SExtractor auto aperture 
may miss a great portion of the flux of faint objects
given the much broader PSFs at the longer IRAC wavelengths.

Some bright sources have $>0.1$ mag differences 
between their IRACLEAN and SExtractor fluxes.
According to the flags provided by SExtractor,
more than 70\% of objects brighter than 50$\mu$Jy are blended
with their neighbors.
In most of these cases, the SExtractor fluxes are brighter than
the IRACLEAN fluxes, consistent with their
SExtractor fluxes being boosted by their neighbors.
In a handful of cases, the large mag differences are 
a consequence of terminating 
IRACLEAN on saturated/bright extended objects
as discussed in Section~\ref{issues}.
We also verify that all the objects with large differences
between their IRACLEAN and SExtractor fluxes are blended
with their neighbors. 
According to Section~\ref{iracmock},
IRACLEAN can provide better flux
estimates for faint sources in the IRAC images,
and thus the majority of scatter in Figure~\ref{irac-comp}
would be due to the limitation of SExtractor.

\subsubsection{Comparison with the FIREWORKS Catalog}\label{comp_fire}
We compared our IRAC fluxes with those from the FIREWORKS catalog
\citep{wuyts2008}, which provides the fluxes for the IRAC GOODS-S data.
\citet{wuyts2008} used the segmentation map generated 
from the VLT ISAAC $K_S$-band
\citep{retzlaff2010} image as a prior 
and convolved the $K_S$-band image with IRAC PSFs for each object.
They then tried to remove the neighbors around a certain object in the
IRAC images, by subtracting their flux-matched convolved images.
By assuming that all the neighbors are well-subtracted for the target,
they then directly put an aperture to measure its flux.
This is an alternative method to minimize the blending issue 
for IRAC flux measurements,
and it is very useful to compare the relative merits of IRACLEAN 
and the method used by \citet{wuyts2008}.

\begin{figure*}
\epsscale{1.0}
\plotone{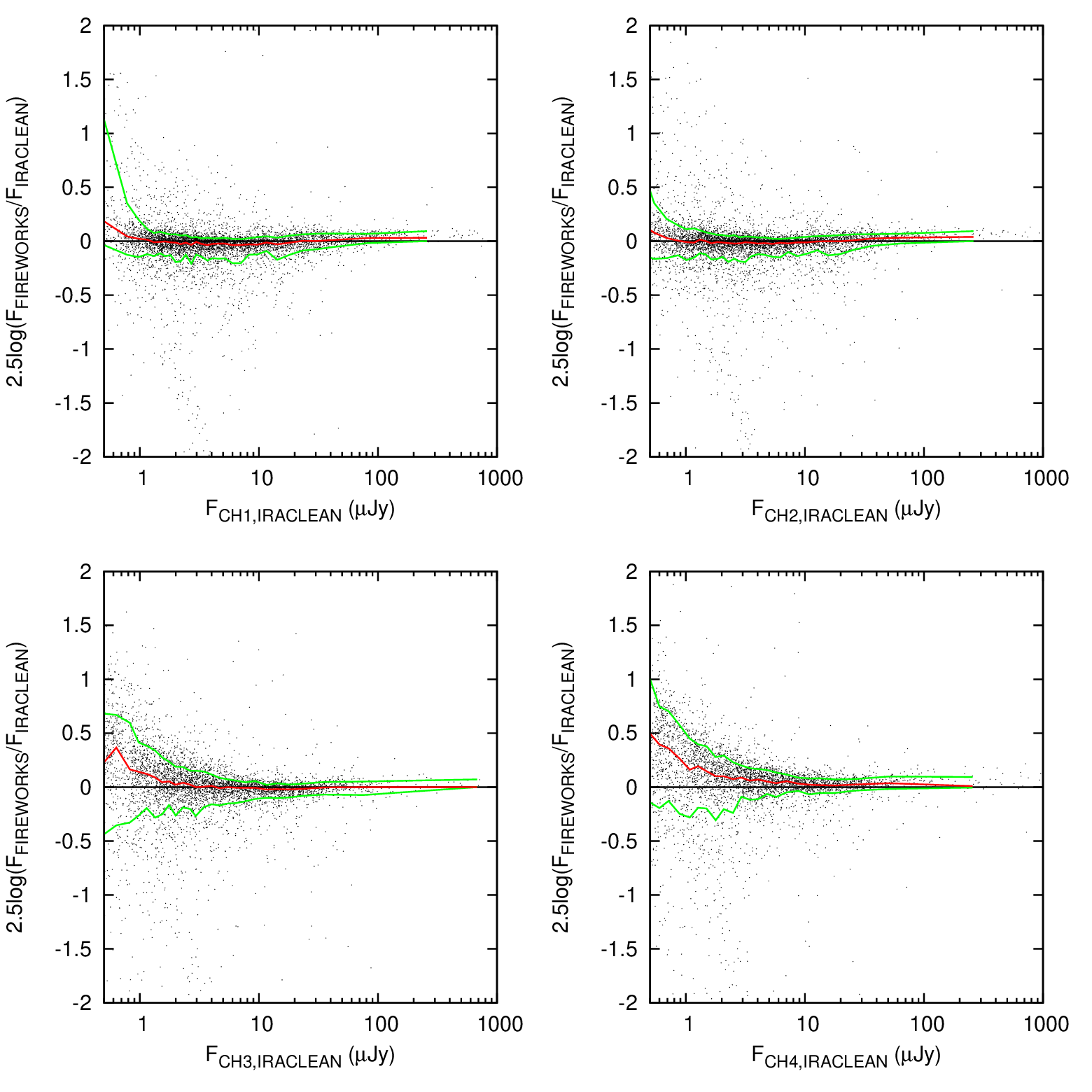}
\caption{IRAC photometry comparison between our catalog (IRACLEAN method) 
and the FIREWORKS catalog.
The red line indicates the running median
while the green lines indicate the upper and lower 68th percentiles.
The majority of objects have similar fluxes in the two catalogs
with an rms of flux ratio dispersions of less than 8\%.
\label{tenis-fireworks} }
\end{figure*}

We show the comparison in Figure~\ref{tenis-fireworks}.
In general the IRAC fluxes between the two catalogs 
are consistent with each other.
The very small dispersions in Figure~\ref{tenis-fireworks},
especially for 3.6 and 4.5 $\mu$m, 
imply that these two entirely independent methods both work well 
in terms of resolving the IRAC blending issue.
There are, however, small differences between the two catalogs.
At the bright end, objects brighter than 30 $\mu$Jy
have $\sim5$\% higher fluxes in the FIREWORKS catalog.
At the faint end, particularly at 5.8 and 8.0 $\mu$m, 
objects also have higher fluxes in the FIREWORKS catalog.
The exact nature of these small differences are unclear to us.
They are unlikely due to under-estimated IRACLEAN fluxes for blended objects,
as we demonstrated in our Monte Carlo simulations (Section~\ref{montecarlo}).
We suspect that they are caused by the unmatched PSFs in \citet{wuyts2008},
since they did not deconvolve their $K_S$ image with the $K_S$ PSF before
the convolution with the IRAC PSFs, 
which may lead to over-subtracting the outer wings of 
the neighboring objects and cause under-estimated local background. 
However, this is hard to verify with the data we have and without
knowing the exact procedure in \citet{wuyts2008}.

Notice that the scatter in the distributions at the faint end
at 3.6 and 4.5 $\mu$m in Figure~\ref{tenis-fireworks} are very small.
This suggests that the reference image (i.e., the ISAAC $K_S$ image)
is not as deep as the IRAC 3.6 and 4.5 $\mu$m images,
so the fluxes of many fainter IRAC objects were not measured and 
included into the FIREWORKS catalog.
Moreover, the faint IRAC objects that are not detected 
in the ISAAC $K_S$ image were not subtracted 
before aperture photometry was applied to their brighter neighbors.
Therefore, the fluxes of their brighter neighbors may be over-estimated.
They may also contribute to the background noise and
make the FIREWORKS detection limits worse.

\subsubsection{Comparison with the SIMPLE Catalog}\label{comp_simple}
\begin{figure*}
\epsscale{1.0}
\plotone{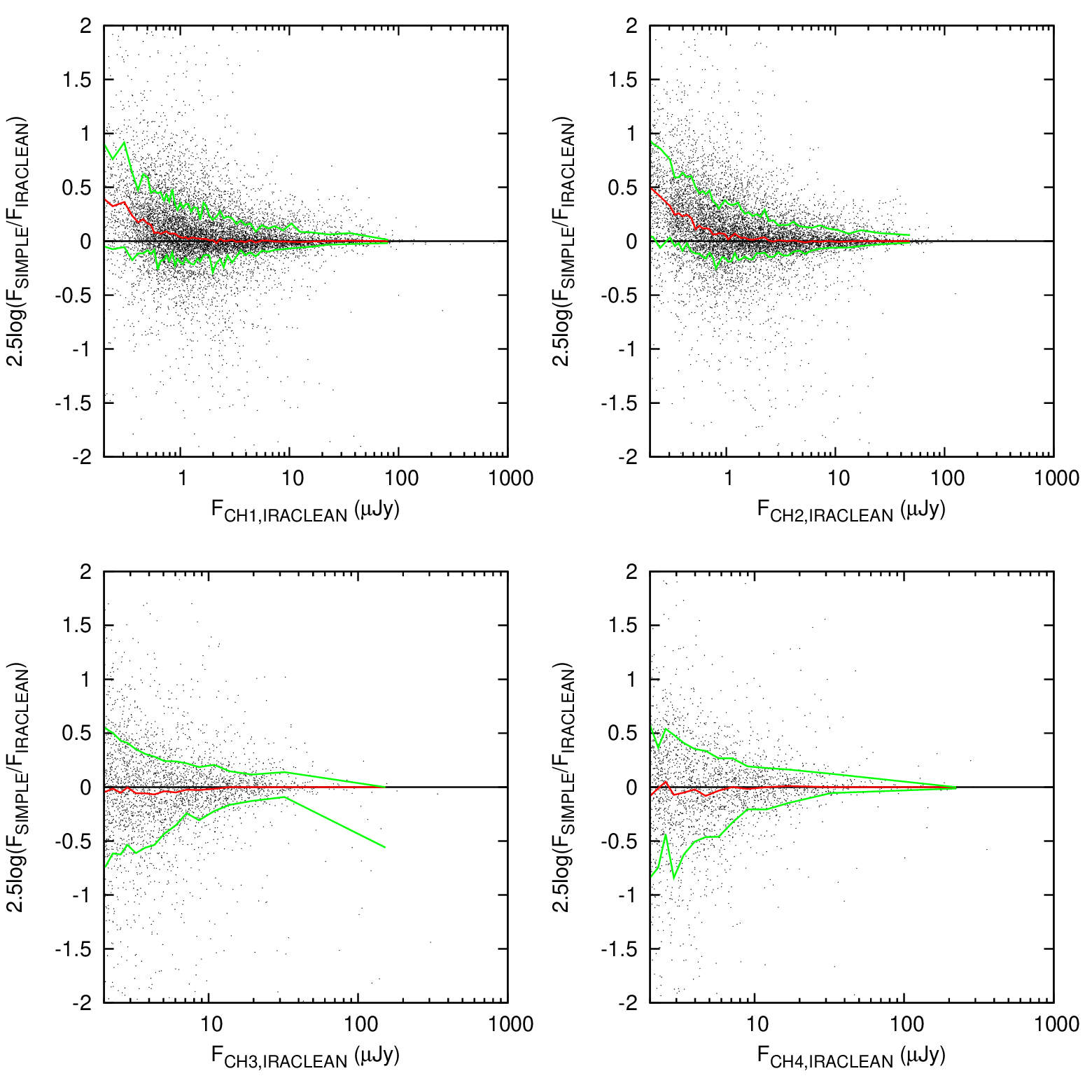}
\caption{Photometry comparison between the IRACLEAN flux and
the SIMPLE catalog.
The red line indicates the running median
while the green lines indicate the upper and lower 68th percentiles.
Only isolated objects are shown as suggested by \citet{damen2011}.
According to this figure, the fluxes of isolated objects
are consistent between both catalogs very well.
\label{tenis-simple} }
\end{figure*}

\citet{damen2011} published a photometric catalog of the SIMPLE data.
They used AUTO aperture in SExtractor to measure fluxes of all objects.
If the AUTO aperture for an object is smaller than $4''$ in diameter,
they replaced the AUTO flux with flux measured 
with a fixed $4''$ diameter aperture.
If objects are blended, their $4''$-aperture fluxes are also used.
They then applied  aperture corrections on all these objects.
\citet{damen2011} excluded blended objects when they compared
their source fluxes with the FIREWORKS catalog, 
claiming that these objects worsen the comparison.
We thus first compare our fluxes with the SIMPLE catalog on
isolated objects according to the blended flag in the SIMPLE v3.0 catalog.
The result is shown in Figure~\ref{tenis-simple} 
and the comparison is fairly good.
We emphasize, however, that more than 70\% of the matched objects are
marked as blended sources in the SIMPLE catalog.
The result shown in Figure~\ref{tenis-simple}
is therefore not a representative comparison between the two catalogs.

\begin{figure*}
\epsscale{1.0}
\plotone{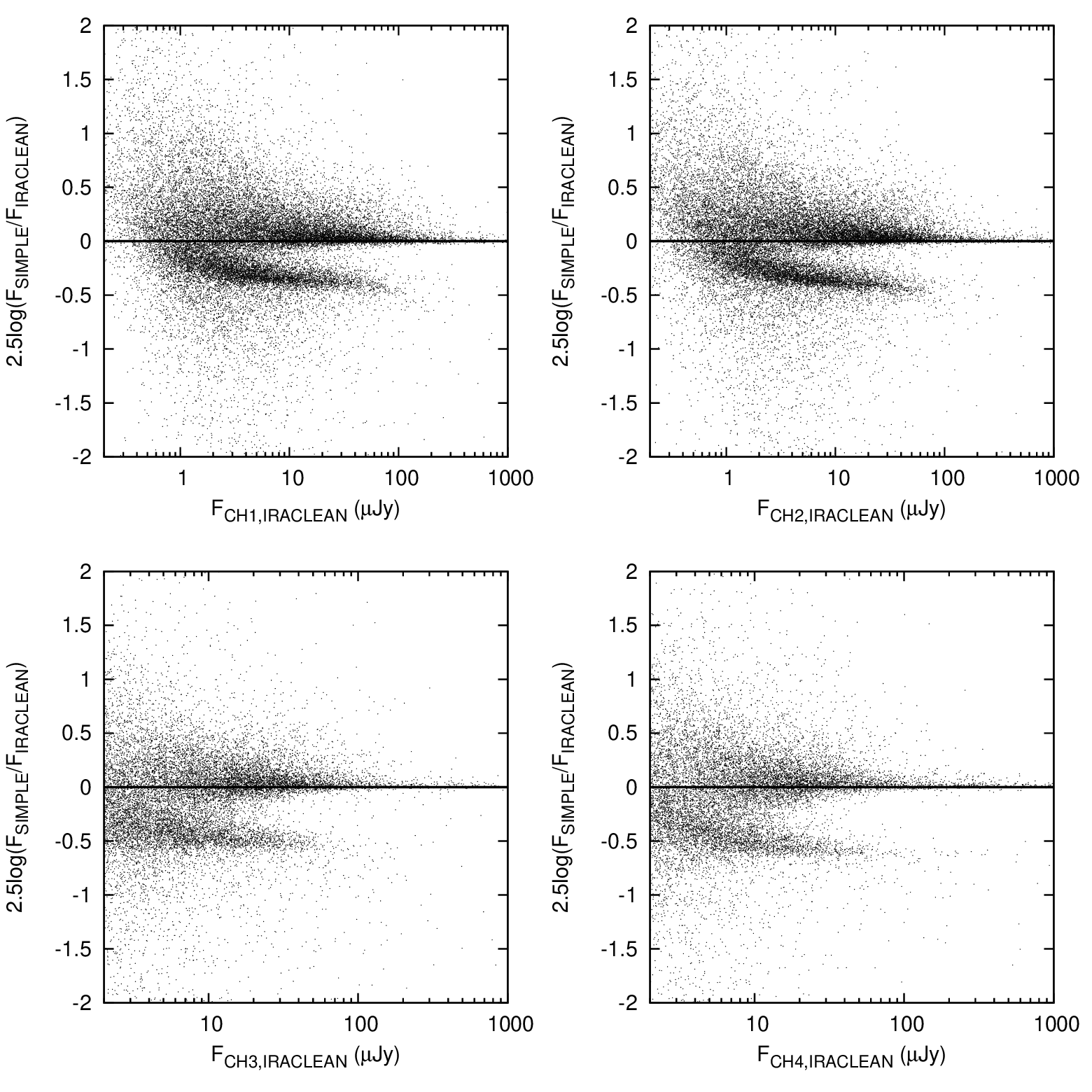}
\caption{Same as Figure~\ref{tenis-simple} but for blended objects.
\label{tenis-simple-blended} }
\end{figure*}

We next compare our IRACLEAN results with the SIMPLE v3.0 catalog
on blended objects.  
The results, shown in Figure~\ref{tenis-simple-blended}, are quite striking.  
There are two distinct sequences in each IRAC band.
The upper sequences contain approximately 66\%, 68\%, 54\% and 51\% 
of all blended objects at 3.6, 4.5, 5.8, and 8.0 $\mu$m, respectively. 
The SIMPLE fluxes of the upper sequences 
are only slightly higher compared with our IRACLEAN fluxes.
The reason for the differences is that
the SIMPLE fluxes of blended sources would be over-estimated
because conventional aperture photometry method was used,
while the IRACLEAN method is designed 
to estimate relatively unbiased fluxes for blended sources.
On the other hand, the SIMPLE fluxes of the lower
sequences are between 30\% to 60\% lower than the IRACLEAN fluxes. 
This very large offset, and the fact that there are two distinct sequences, 
cannot be explained by any systematic effects that we are aware of.  
We looked at the objects in the lower sequence objects in the IRAC images, 
but did not find anything different about 
these objects compared with those in the upper sequences.
We remind the reader at this point that
both IRACLEAN and the method used by \citet{damen2011}
to derive fluxes of indiviaul objects are based on the same SIMPLE IRAC images,
yet we do not see such second sequences when we compare our results
with XCONV, SExtractor, and FIREWORKS.  
The exhaustive comparisons 
that we have made with all relevant existing catalogs suggest that
the lower sequences are previously unknown systematic effects 
in the SIMPLE v3.0 catalog.   
We are therefore concerned about the conclusions reached in studies 
that have been made based on the SIMPLE v3.0 catalog.

\section{Combined NIR and IRAC Catalog}\label{mastercatalog}
We combined the TENIS WIRCam photometry and the SIMPLE IRACLEAN photometry.
We also removed objects with fluxes less than 3$\sigma$ 
in both the $J$ and $K_S$ bands.
The final TENIS WIRCam and IRAC catalog includes 62,326 objects.
We did not apply Galactic extinction correction to this catalog.
We also include the XCONV colors 
since they are good color references for isolated bright sources.
Table~\ref{catalog_statistics} shows the statistics of this catalog
which has the following format:

\vspace{3mm}
$(1)$: Object ID.

\vspace{3mm}
$(2)-(3)$: R.A. and Decl.

\vspace{3mm}
$(4)-(15)$: Flux and flux error measurements (in $\mu$Jy) for
$J$, $K_S$, IRAC $3.6\mu{m}$, $4.5\mu{m}$, $5.8\mu{m}$,
and $8.0\mu{m}$.

\vspace{3mm}
$(16)-(19)$: XCONV $K_S$ - IRAC colors (in AB magnitude).

\begin{deluxetable}{lcc}
\tabletypesize{\scriptsize}
\tablecolumns{3}
\tablewidth{0pt}
\tablecaption{Number of Objects and Detection Limits 
in the Multiband Catalog\label{catalog_statistics}}
\tablehead{
\colhead{Waveband} & \colhead{Detections} & \colhead{Depth ($\mu$Jy)}}
\startdata
$J$ & 53722 & 0.115\\
$K_S$ & 57492 & 0.199\\
$3.6\mu{m}$ & 53801 & 0.083\\
$4.5\mu{m}$ & 49667 & 0.105\\
$5.8\mu{m}$ & 28284 & 0.541\\
$8.0\mu{m}$ & 22418 & 0.669\\
\enddata
\tablecomments{
Detections are the number of objects with S/N $>$ 3.
Depths are the medium values of $1\sigma$ flux errors 
for objects with S/N $>$ 3. }
\end{deluxetable}

\subsection{Depths}\label{depth}
The sensitivity of our catalog are by far 
the highest in all the wavebands from $J$ to 8.0 $\mu$m
among existing observations in the ECDFS and GOODS-S.
In the ECDFS region, 
the median $5\sigma$ limits among all detected objects in our catalog
are 24.50, 23.91, 24.85, 24.60, 22.82, and 22.59 mag
for $J$, $K_S$, 3.6, 4.5, 5.8, and 8.0 $\mu$m, respectively.
These $J$ and $K_S$ limiting magnitudes are much deeper than 
those of the MUSYC survey \citep{cardamone2010}.  
As shown in Section~\ref{isaaccomp},
our $K_S$ depth in the GOODS-S region is also $\sim0.5$ mag deeper than 
the VLT ISAAC data published in \citet{retzlaff2010}.  
The same is also true for our $J$ depth,
as compared to the VLT ISAAC $J$ depth in \citet{grazian2006}.

We found that our IRAC limiting magnitudes are much deeper 
than those provided in \citet{damen2011}, 
which are 23.8, 23.6, 21.9, and 21.7 mag
at 3.6, 4.5, 5.8, and 8.0 $\mu$m, repectively.
Since our work and that of Damen et al.\ are based on the same IRAC dataset, 
it is thus important to examine whether our sensitivities are reasonable.
We first compare the residual noises in our IRACLEAN images with those 
provided by $Spitzer$ Sensitivity Performance Estimation Tool (SENS-PET).
The detection limit estimated by SENS-PET
is derived using a 10-pixel radius aperture ($\sim24''$ in diameter) 
for point-like sources. 
We therefore measured the rms of the background noise 
in our IRACLEAN residual images with a $24''$ diameter aperture. 
They are 0.28, 0.40, 1.74, and 1.74 $\mu$Jy 
for 3.6, 4.5, 5.8, and 8.0$\mu$m, respectively,
and where the average integration time 
in the SIMPLE IRAC images is $\sim1.5$ hr.
According to the SENS-PET, 
1 hr of integration will provide $1\sigma$ sensitivities of
0.191, 0.277, 1.56, and 1.67 $\mu$Jy 
for 3.6, 4.5, 5.8, and 8.0$\mu$m, respectively. 
Our sensitivities are worse,  
which is expected since the sensitivities quoted 
by the SENS-PET correspond to the ideal cases.
Next, we convert the SENS-PET sensitivities to 
our IRACLEAN $2\farcs7 \times 2\farcs7$ aperture, 
by assuming that the photometric error scales 
with square-root of the aperture area.
After taking into account the aperture corrections (Section~\ref{IRACfluxerr}),
we obtained 5 $\sigma$ limiting magnitudes of 25.51, 25.12, 23.12, and 22.93 
for 3.6, 4.5, 5.8, and 8.0 $\mu$m, respectively. 
Our actual limiting magnitudes are again shallower.

The above comparisons show that our sensitivities 
are better than what had been previously achieved on the SIMPLE IRAC data, 
but do not exceed what can be achieved in ideal cases.
The differences in the detection limits between our catalog
and that of  \citet{damen2011} are most likely caused by
confusion effects in the IRAC images. 
With our $J$+$K_S$ prior image that is nearly as deep
as the IRAC images, IRACLEAN is much less affected by faint undetected sources
as well as the PSF wings of nearby bright objects.
This allows us to estimate fluxes for faint sources more accurately, and
thus pushing the detection limits closer to the ideal values.

\subsection{Spurious sources}\label{spurious}
As we mentioned in the beginning of this section, 
we cleaned the TENIS multi-wavelength catalog
by removing objects with low S/N.
Although this step should have eliminated most spurious objects,
the catalog could still be significantly contaminated.
We therefore investigated the spurious fraction in our catalog.
First, we inverted all the images (i.e., making negative images)
including the $J$, $K_S$, $J$+$K_S$, and the four IRAC images.
We then repeated identical steps used to 
generate the multi-wavelength catalog on these inverted images.
The final catalog for the inverted images contains 21578 objects,
which suggests that the spurious fraction of the TENIS multi-wavelength catalog
is unreasonably high ($\sim35\%$). 
We checked the inverted $J$+$K_S$ image and the ``sources'', 
and we found that most of the ``sources'' correspond to negative holes 
produced by the crosstalk removal procedure of the WIRCam reduction pipeline. 
(The crosstalk removal does not generate positive features.) 
This is similar to the case in \citet{wang2010}. 
If we just use the relatively crosstalk-free regions 
to calculate the spurious fraction, the value dramatically decreases to 6\%. 
We note that the value of 6\% is an upper limit 
since fainter objects still can produce low-level crosstalk. 

\begin{figure}
\epsscale{1.0}
\plotone{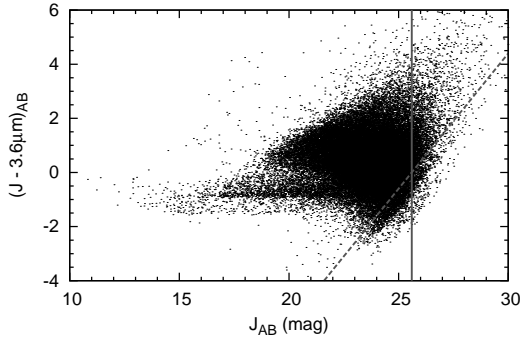}
\caption{
$J$ and 3.6$\mu$m color-magnitude diagram. 
The gray solid-line indicates the 5$\sigma$ limiting magnitude 
for point sources in $J$,
and the gray dashed-line indicates the 5$\sigma$ limit of $J$--$3.6\mu$m color
for point sources.
The blue galaxies with $J$ -- 3.6$\mu$m $< -0.4$ start disappearing at $J > 25$ 
(the bottom-right corner), 
which implies that many objects in the TENIS catalog with S/N $<$ 3 at 3.6$\mu$m 
are real J-detected objects.
\label{JCH1} }
\end{figure}

A second test on the spurious fraction is objects detected 
by both WIRCam and IRAC. 
According to Table~\ref{catalog_statistics}, 
there are 53801 objects detected at 3.6$\mu$m with S/N $>$ 3, 
suggesting a spurious fraction of 14\%. 
However, this is also an upper limit. 
The $J$ and 3.6$\mu$m color-magnitude diagram in Figure~\ref{JCH1} shows that 
blue galaxies with $J$ -- 3.6$\mu$m $<$ -0.4 start disappearing at $J > 25$. 
This indicates that many IRAC undetected but $J$ detected objects are real. 
Therefore, the spurious fraction must be much less than 14\%. 
Based on the above two tests, 
we conclude that the spurious fraction of the TENIS catalog is less than 6\%.

We do not attempt to quantify the completeness
of our catalog, which is nontrivial given the complex nature of IRACLEAN and
the fact that we detect sources from a $J$+$K_S$ image. 
Readers interested in using our catalog
and wishing to know the completeness at a given flux level should run 
their own source extraction and selection, quantify the completeness, 
and then extract photometry from our multiwavelength catalog.

\section{Summary}\label{summary}
We present an ultra-deep $J$ and $K_S$ dataset
covering a $30\arcmin \times30\arcmin$ area in the ECDFS,
as part of our Taiwan ECDFS Near-Infrared Survey (TENIS).
The median $5\sigma$ limiting magnitudes for all objects
reach 24.5 and 23.9 mag (AB) for $J$ and $K_S$, respectively.
In the inner 400 arcmin$^2$ region of the images
where the sensitivity is more uniform,
objects as faint as 25.6 and 25.0 mag are detected at 5 $\sigma$.
In our final catalog, we detect objects in a $J$+$K_S$ image in order to
achieve higher completeness.
We also developed a novel deconvolution technique (IRACLEAN)
to accurately estimate the IRAC fluxes 
for all the $J$+$K_S$ detected objects in the ECDFS, using
our $J$+$K$ image as a prior.
With simple Monte-Carlo simulations and 
comparison against the XCONV technique, we showed that
IRACLEAN is able to correctly recover IRAC fluxes for most objects.
We also compared IRACLEAN fluxes with fluxes directly measured by SExtractor,
and with the FIRWORKS and SIMPLE catalogs. We found that IRACLEAN results
are superior in many cases, and our IRACLEAN results provide 
by far the deepest IRAC catalog in the ECDFS region.
This $J$+$K_S$ detected catalog consists of
flux measurements for $J$, $K_S$,
IRAC 3.6, 4.5, 5.8, and 8.0 $\mu$m, 
and XCONV $K_S$--IRAC colors for all four IRAC bands.
We publicly release the data products of this work, 
including the $J$ and $K_S$ images, 
and the $J$+$K_S$ selected multiwavelength catalog.

\acknowledgments
We thank the referee for comments that greatly improve the manuscript.
We are grateful to the Hawaiian group led by Lennox Cowie
for contributing their WIRCam $K_S$-band data,
and to the CFHT staff for help in obtaining the data.
This paper is based on observations obtained with WIRCam, 
a joint project of CFHT, Taiwan, Korea, Canada, France, 
and the Canada-France-Hawaii Telescope (CFHT) 
which is operated by the National Research Council (NRC) of Canada, 
the Institute National des Sciences de l'Univers of 
the Centre National de la Recherche Scientifique of France, 
and the University of Hawaii.
Access to the CFHT was made possible by the Ministry of Education,
the National Science Council of Taiwan as part of the Cosmology and 
Particle Astrophysics (CosPA) program,
the Academia Sinica, Institute of Astronomy and Astrophysics,
and the National Tsing Hua University, Taiwan.
We gratefully acknowledge support from the National Science Council of Taiwan
grant 98-2112-M-001-003-MY2 (W.H.W.), 99-2112-M-001-012-MY3 (W.H.W).


\begin{thebibliography}{}

\bibitem[Anderson \& King(2003)]{anderson2003}
Anderson, J., \& King, I.\ R.\ 2003, \pasp, 115, 113

\bibitem[Bertin \& Arnouts~(1996)]{ba1996}
Bertin, E. \& Arnouts, S.~1996, A\&AS, 117, 393

\bibitem[Caldwell et al.~(2008)]{caldwell2008}
Caldwell, J. A. R., et al.~2008, \apjs, 174, 136

\bibitem[Capak et al.~(2007)]{capak2007}
Capak, P., et al.~2007, \apjs, 172, 99

\bibitem[Cardamone et al.~(2010)]{cardamone2010}
Cardamone, C. N., can Dokkum, P. G., Urry, C. M., et al.~2010, \apjs, 189, 270

\bibitem[Cohen et al.~(2003)]{cohen2003}
Cohen, M., Wheaton, Wm. A., \& Megeath, S. T.~2003, \aj, 126, 1090 

\bibitem[Dahlen et al.~(2010)]{dahlen2010}
Dahlen, T., et al.~2010, \apj, 724, 425

\bibitem[Damen et al.~(2011)]{damen2011}
Damen, M., et al.~2011, \apj, 727, 1

\bibitem[Elston, Rieke, \& Rieke~(1988)]{err1988}
Elston, R., Rieke, G. H., \& Rieke, M. J.~1988, \apj, 331, L7

\bibitem[Ford et al.~(2003)]{ford2003}
Ford, H. C., et al.~2003, Proc. SPIE, 4854, 81

\bibitem[Franx et al.(2003)]{franx2003}
Franx, M., Labb\'{e}, I., Rudnick, G., et al.\ 2003, \apjl, 587, L79

\bibitem[Gawiser et al.~(2006)]{gawiser2006}
Gawiser, E., et al.~2006, \apj, 642, L13

\bibitem[Giavalisco et al.(2004)]{giavalisco2004}
Giavalisco, M., Ferguson, H.\ C., Koekemoer, A.\ M., et al. 2004, \apjl, 600, L93

\bibitem[Grazian et al.~(2006)]{grazian2006}
Grazian, A., et al.~2006, A\&A, 449, 951

\bibitem[Grogin et al.~(2011)]{grogin2011}
Grogin, N. A., et al.~2011, \apjs, 197, 35

\bibitem[Guo et al.~(2012)]{guo2012}
Guo, Y., et al.~2012, \apj, 749, 149

\bibitem[H\"{o}gbom et al.~(1974)]{hogbom1974}
H\"{o}gbom, J. A.~1974, A\&A Suppl., 15, 417

\bibitem[Hsieh et al.~(2012)]{hsieh2012}
Hsieh, B.C., Wang, W.-H., Yan, H., Lin, L., Karoji, H.,
Lim, J., Ho, P. T. P., \& Tsai, C. W.~2012, \apj, 749, 88

\bibitem[Koekemoer et al.~(2011)]{koekemoer2011}
Koekemoer, A. M., et al.~2011, \apjs, 197, 36

\bibitem[Kurucz~(1993)]{kurucz1993}
Kurucz, R. L.~1993, Kurucz CD-ROM 19, 
ATLAS9 Stellar Atmosphere Programs and 2 kms Grid (Cambridge: SAO)

\bibitem[Laidler et al.~(2007)]{laidler2007}
Laidler, V. G., et al.~2007, \pasp, 119, 1325

\bibitem[McLure et al.~(2011)]{mclure2011}
McLure, R. J., et al.~2011, arXiv1102.4881

\bibitem[Ouchi et al.~(2009)]{ouchi2009}
Ouchi, M., et al.~2009, \apj, 706, 1136

\bibitem[Puget et al.~(2004)]{puget2004}
Puget, P., et al.~2004, SPIE, 5492, 978

\bibitem[Retzlaff et al.~(2010)]{retzlaff2010}
Retzlaﬀ, J., Rosati, P., Dickinson, M., Vandame, B., Rité, C., 
Nonino, M., Cesarsky, C., and the GOODS Team~2010, A\&A, 511, 50

\bibitem[Rix et al.~(2004)]{rix2004}
Rix, H. -W., et al.~2004, \apjs, 152, 163

\bibitem[Skrutskie et al.~(2006)]{skrutskie2006}
Skrutskie, M. F., et al.~2006, \aj, 131, 1163

\bibitem[Sawicki~(2002)]{sawicki2002}
Sawicki, M.~2002, \aj, 124, 3050

\bibitem[Simpson \& Eisenhardt~(1999)]{se1999}
Simpson, C., \& Eisenhardt, P.~1999, \pasp, 111, 691

\bibitem[Taylor et al.~(2009)]{taylor2009}
Taylor, E. N.~2009, \apjs, 183, 295

\bibitem[Wang et al.~(2010)]{wang2010}
Wang, Wei-Hao, Cowie, Lennox L., Barger, Amy J., Keenan, Ryan C., 
Ting, Hsiao-Chiang~2010, \apjs, 187, 251

\bibitem[Wang, Cowie, \& Barger~(2012)]{wang2012a}
Wang, Wei-Hao, Cowie, Lennox L., \& Barger, Amy J.~2012, \apj, 744, 155

\bibitem[Wolf et al.~(2001)]{wolf2001}
Wolf, C., Dye, S., Kleinheinrich, M., Meisenheimer, K.,
Rix, H.-W., Wisotzki, L.~2001, A\&A, 377, 442

\bibitem[Wuyts et al.~(2008)]{wuyts2008}
Wuyts, S., Labb\'{e}, I., F\"{o}rster-Schreiber, N. M., Franx, M., Rudnick, G.,
Brammer, G. B., \& van Dokkum, P. G.~2008, \apj, 682, 985

\bibitem[Yan et al.(2004)]{yan2004}
Yan, H., Dickson, M., Eisenhardt, P.\ R.\ M., et al.\ 2004, \apj, 616, 63

\bibitem[Yan et al.~(2011)]{yan2011}
Yan, H., et al.~2011, \apj, 728, 22

\end{thebibliography}
\end{document}